\documentclass[usenatbib,usegraphicx]{mn2e}
\usepackage{bethmacros}
\usepackage{epsfig,psfig,phase14}
\begin{document}

\title[Forming Disk Galaxies]{Forming Disk Galaxies in $\Lambda$CDM Simulations}

\author[Governato et al.]
{
\parbox[t]{\textwidth}{
\vspace{-1.0cm}
F. Governato$^{1,2}$, 
B. Willman $^{3}$,
L. Mayer$^{4}$,
A. Brooks$^{1}$,
G.Stinson$^{1}$,
O.Valenzuela$^{1}$,
J.Wadsley$^{5}$, 
T.Quinn$^{1}$
}
\vspace*{6pt} \\
$^1$Department of Astronomy, University of Washington, Box 351580, Seattle, WA 98195, USA \\
$^2$INAF, Osservatorio Astronomico di Brera, via Brera 29, 20121, Milano, Italy  \\
$^3$NYU, Department of Physics,  4 Washington Place, New York, NY 10003  \\
$^4$ETH, Ramistrasse 101, CH-8092 Zurich.\\
$^5$Department of Physics and Astronomy, McMaster University, Hamilton, Ontario L88 4M1, Canada
\vspace*{-0.5cm}}

\maketitle


\begin{abstract}

We used fully cosmological, high resolution N-body + SPH simulations
to follow the formation of disk galaxies with rotational velocities
between 135 and 270 km/sec in a $\Lambda$CDM universe.  The
simulations include gas cooling, star formation, the effects of a
uniform UV background and a physically motivated description of
feedback from supernovae. The host dark matter halos have a spin and
last major merger redshift typical of galaxy sized halos as measured
in recent large scale N--Body simulations. The simulated galaxies form
rotationally supported disks with realistic exponential scale lengths
and fall on both the I-band and baryonic Tully Fisher relations.  An
extended stellar disk forms inside the Milky Way sized halo
immediately after the last major merger. The combination of UV
background and SN feedback drastically reduces the number of visible
satellites orbiting inside a Milky Way sized halo, bringing it in fair
agreement with observations.  Our simulations predict that the average
age of a primary galaxy's stellar population decreases with mass,
because feedback delays star formation in less massive
galaxies. Galaxies have stellar masses and current star formation
rates as a function of total mass that are in good agreement with
observational data.  We discuss how both high mass and force
resolution and a realistic description of star formation and feedback
are important ingredients to match the observed properties of
galaxies.

\end{abstract}




\section{Introduction}
\label{intro}

In a universe dominated by Cold Dark Matter (CDM) and a cosmological
constant \citep{blumenthal84,bbks,spergel03,peeblesratra03}, galaxy
formation and evolution is a complex combination of hierarchical
clustering, gas dissipation, merging and secular evolution. While
gravity drives the bottom--up assembly of cosmic structures
\citep{davis85}, gas cools at the centers of dark matter halos and
acquires angular momentum through tidal torques from nearby structures
\citep{fall80,fall83}. Cool gas eventually fragments due to Jeans
instabilities and forms stars \citep{whiterees78,peebles69}.
Gravitational interactions between galaxies can instigate star
formation and transform disks into spheroids \citep{toomre72} and may
dominate galaxy evolution at early epochs especially in dense
environments \citep{frenk85}.  Secular processes, such as external gas
accretion or gas displacement due to bars, may play a more important
role in galaxy evolution at later epochs
\citep{valenzuela03,debattista04,kormendy04}. Galaxy formation in this
framework is a continuous, ongoing process where the observable
properties of galaxies are a function of their merging histories,
masses and environments
(e.g. \citealp{shen03,blanton03,brinchmann04,bundy05}).

N--Body/gasdynamical simulations have become the primary tools with
which to model galaxy formation in a cosmological context.  They are
necessary to follow the evolution of the internal structure of
galaxies as well as the complex interplay between baryon cooling and
feedback \citep{lakecarlberg88, katz92,navarro94,barneshernquist96,
quinn96,haehnelt98,maclow99,cen99,thacker00,navarrosteinmetz00,tittley01,
sommer03,governato04}. However, numerical simulations have had
difficulties reproducing the properties of real galaxies. Early work
reported a catastrophic loss of angular momentum in the baryonic
component of simulated galaxies, leading to the formation of galaxies
 dominated by a central concentration of cold baryons
\citep{navarro94}.  As predicted by early theoretical models
\citep{whiterees78, whitefrenk91}, simulations without strong energy
feedback from stellar processes have also suffered from overly
efficient star formation. This causes galaxies to form most of their
stellar component as soon as gas is able to collapse and cool within
small halos, well before the assembly of the main progenitor
\citep{balogh01}. Dense clumps of stars and cold gas might then
spiral rapidly at the center of the main galaxy due to dynamical
friction \citep{navarrosteinmetz97}.  These shortcomings resulted
in compact galaxies with a central excess of old stars and offset from the
observed Tully-Fisher relation (\citealt{giovanelli97a}). The offset
was $\sim$ 1.5 mag for an $\Omega_0$ = 1 CDM cosmology
\citep{navarrosteinmetz00,eke01} and $\sim$ 0.5 mag in a high $\sigma_8$ $\Lambda$CDM \citep{eke01}.

More recently, attention has been
drawn to the ``missing galaxy problem''. Dark matter only simulations
predict over an order of magnitude more subhalos around Milky Way-like
galaxies than the number of dwarfs observed around the Milky Way and
M31 \citep{klypin99,moore99, willman04}.  Observations also show that
more massive late-type galaxies have older stellar populations than
less massive late-type galaxies
\citep{macarthur04,ferreras05,gallazzi05}, in possible contradiction
with the fact that in CDM scenarios less massive halos assemble (on
average) first.

The angular momentum and missing galaxy problems have often been
linked to two major stumbling blocks for numerical galaxy formation:
i) insufficient numerical resolution and ii) inaccurate treatment of
feedback due to star formation.  Only the most recent gas dynamical
simulations have achieved spatial resolution sufficient to resolve the
disk scale lengths of disks.  \citet{governato04} and
\citet{kaufmann06} have shown that poor mass and spatial resolution
might lead to significant numerical angular momentum loss in baryonic
disks embedded in dark matter halos. These works suggested that at
least 10$^5$ DM particles (and equivalent numbers in gas and star
particles) within the virial radius as well as force resolution of
$\leq 1$ kpc are required to faithfully simulate the formation and
dynamical evolution of disk systems over cosmic times.  Recent
simulations \citep{brook04,abadi03a,robertson04,okamoto05} that used a
large number of particles and a high force resolution have indeed
formed galaxies with extended stellar disks. However, most of the
galaxies formed in most of these simulations also had a massive
spheroidal component or disks only partially supported by rotation.

 The impossibility of directly resolving the scales at which star
  formation and feedback happen (a few pc) makes it necessary to
  develop simplified models to describe star formation and subsequent
  energy feedback from supernovae (SNe) at galactic scales (0.1 -- 1
  kpc) \citep{efstathiou00,silk01,wada01,ferrara02,krumholz05,Slyz05}.
  In early simulations, thermal energy that was simply added to gas
  surrounding star forming regions was quickly radiated away,
  resulting in overcooled gas \citep{katz92}.  However, heated,
  diffuse baryons resulting from a proper treatment of feedback may be
  less susceptible to angular momentum transfer to halo particles
  \citep{mo96,eke00,maller02,donghia06}. To address this shortcoming,
  \citet{gerritsen97}, \citet{yepes97} and \citet{thacker00} modeled
  the pressure support stimulated by gas turbulence by shutting off
  cooling on timescales of a few million years, and reported the
  formation of larger disks.  \citet{brook04} used a scheme similar to
  that of \citet{thacker00} and also found it produced reasonably
  realistic galaxy disks when applied to simplified cosmological
  initial conditions.  \citet{robertson04} used an approach similar to
  that of \citet{springel00} and found that the pressure support
  produced by their multiphase treatment of the ISM was an important
  factor in the formation of large disks.  Supernova feedback and the
  UV background, being able to reduce gas retention and accretion in
  halos with low virial temperature, may also solve the problem of the
  overabundance of satellites
  \citep{quinn96,gnedin00,benson02,dekel03,monaco04,kravtsov04}.
  Other recent papers have investigated the roles of a top heavy
  IMF \citep{okamoto05} and of massive galaxy outflows driven by QSO
  activity \citep{binney01,granato04,dimatteo05} in regulating galaxy
  formation.

In this paper, we study the effect of feedback on the structure and
satellites of disk galaxies formed within cosmological halos spanning
a significant mass range (from 10$^{12}$ M$_{\odot}$ down to
10$^8$). We improve over previous work in two ways.  First, the mass
and spatial resolution of these simulations are sufficient to resolve
(i) the structure of present day disks without being significantly
limited by resolution and (ii) the sub-halo population for each galaxy
in our sample down to circular velocities of about 20\% of their
parent halo allowing us study the basic properties of galactic
satellites.  Second, we use a revised implementation of the star
formation algorithm introduced by \citet{katz92} combined with the
feedback recipe introduced by \citet{gerritsen97} and then studied by
\citet{thacker00}. Details of our implementation can be found in
\citet{stinson06} (hereafter S06). In this paper we
calibrate the free parameters of our algorithm to accurately describe
the star formation in isolated galaxy models of present day galaxies.
We then apply this algorithm to cosmological simulations of individual
high resolution galaxies spanning a decade in mass and analyze their
properties.

\smallskip

We focus our analysis in this paper on three fundamental properties of
present day galaxies:

\begin{itemize}
\item{ The abundance and luminosities of galactic satellites.}

\item{ The Tully--Fisher and baryonic Tully--Fisher relations
\citep{giovanelli97a,mcgaugh05}}.
 
\item{ Global star formation histories (SFHs) and z = 0 star formation
rates.}
\end{itemize}

In \S2, we briefly describe the code and the star formation algorithm.
A detailed description of our star formation and feedback algorithm
with additional tests to those presented here is presented in S06.  In
\S3, we summarize the cosmological runs performed with a range of
feedback algorithms, and in \S4-9 we discuss and describe in detail
the properties of our simulated galaxies.  We plan to explore other
aspects of both low and high redshift galaxy formation in future
papers.

\section{The Code and the Star Formation Algorithm}
\label{sec:simulation}
\subsection{GASOLINE}

We used GASOLINE \citep{wadsley04}, a smooth particle hydrodynamic
(SPH), parallel treecode that is both spatially and temporally
adaptive with individual time steps.  The version of GASOLINE we used
implemented i) Compton and radiative cooling, ii) star formation and a
supernova feedback as described in detail in S06 and \citet{katz96},
and iii) a UV background following an updated version of
\citet{haardtmadau96} (Haardt 2005, private communication) starting at
$z = 6$.  The opening angle, $\theta$, was 0.55 until z = 2 and 0.75
thereafter, and the time-stepping criterion, $\eta$, was 0.2, as in
\citet{diemand04}.  The minimum smoothing length allowed is equal to
0.1 times the gravitational softening. Exactly 32 neighbors are used
for SPH calculations.  An ideal gas of primordial composition is
assumed.  We calculate the cooling/heating rate and ionization state
of each particle by assuming collisional ionization equilibrium and
the presence of the time dependent, but uniform UV background.  The
code treats artificial viscosity as suggested in
\citet{Balsara97}. The energy equation is computed asymmetrically
\citep{springel02,evrard88,monaghan92}. This approach avoids the
energy conservation and negative energy problems of the arithmetic and
geometric implementations and converges to the high resolution answer
as well as other recently proposed methods
\citep{benz90,springel02,wadsley04}.

\subsection{The Star Formation and SN Feedback Algorithm}

 Our star formation (SF) algorithm can be broken down into three main
parts: (i) identifying the star forming regions, (ii) forming stars
and (iii) treating stellar evolution including such effects as mass
loss, SN winds and metal enrichment. The algorithm has only three
free parameters: $c{\star}$, $\epsilon$SN, and $\beta$. We define
these parameters and summarize the main features of the algorithm
below.  This algorithm is described in detail in S06.

\subsubsection{Criteria for Star Formation}

The criteria for a gas particle to become eligible for star formation
are: (i) Its temperature is colder than 30,000 K; (ii) Local gas density
is higher than 0.1 cm$^{-3}$; (iii) The gas  particle is part of a
converging flow measured over the 32 nearest neighbors.  We do not
require the criterion that gas particles are Jeans unstable
\citep{kawata06}. A minimum gas overdensity $\delta \rho$/$\rho$ is
also required to avoid spurious star formation at very high z.  We
base the rate at which gas is converted into stars,
$d\rho_{\star}/{dt}$, on the relation

\begin{equation}
 \frac{d\rho_{\star}}{dt} =  \rho_{gas}^{3/2}
\end{equation}

\noindent where $\rho$ represents the volume density.  Using the fact
that dynamical time, $t_{dyn} \propto \rho^{-1/2}$, we express this as

\begin{equation}
\frac{d\rho_{\star}}{dt} = c{\star} \frac{\rho_{gas}}{t_{dyn}}
\end{equation}

\noindent where we have introduced a constant efficiency factor
 $c{\star}$ to enable us to calibrate the star formation algorithm to
 match star formation rates observed for z = 0 galaxies. The mass of
 star particles formed is fixed to 30\% of its parent gas particle
 initial mass \citep{okamoto05}.  Once the particle passes the above
 criteria, to implement Equation (1) in a discrete system we assign a
 probability $p$ that a star will actually be spawned from its parent
 gas particle:

\begin{equation}
p = \frac{M_{SF}}{M_{GP}}(1-e^{-\frac{c{\star} \Delta t}{t_{form}}})
\end{equation}

\noindent where we have introduced $M_{SF}$, the spawning mass for
star particles, $M_{GP}$, the mass of the gas particle that is
creating the star, $\Delta{t}$, the star formation timescale (one Myr
in all of the simulations described in this paper) and $t_{form}$,
which is either the dynamical time or the cooling time, whichever is
longer.  Gas particles in dense regions with shorter dynamical times
will form stars at a higher probability.

\subsubsection{Feedback and Metal Enrichment From Supernovae}

Our implementation of feedback qualitatively follows the algorithm
implemented by \citet{thacker00}.  We assume that the energy released
into the ISM turns into turbulent motions (at unresolved scales) and
is partially dissipated, preventing the gas from cooling and forming
stars.  We determine the number of Supernova type Ia \& II that occur
during each time step from \citet{raiteri96} using a Miller-Scalo IMF
and the stellar lifetimes of stars.  We then multiply the number of
both SN types that explode by a fraction of the canonical $10^{51}$
ergs/SN times a fixed efficiency term ($\epsilon$SN), and distribute
that energy to the surrounding gas particles.  In \S2.3.2, we explore
efficiency values in the range 0.1 to 0.6.  At an efficiency
$\epsilon$SN $=$ 0.1 and with the adopted IMF, $7.65\times10^{47}$
ergs of energy are deposited into the surrounding gas for every one
$M_{\odot}$ of star formed.  Energy is distributed using the smoothing
kernel over the 32 nearest neighbor particles. In our algorithm the
time scale for the cooling shutoff and the amount of mass affected are
physically motivated and chosen following typical values from the
\citet{mckee77} and \citet{ostrikermckee88} blast wave model. When
energy injection comes from a SNII event we disable the radiative
cooling for 30 million years in a number of the nearest neighbor
particles that satisfy:

\begin{equation}
\beta M_{SNII} > \frac{4 \pi r^3}{3} \rho_{ave}
\end{equation}

\begin{table*}
\centering
\begin{tabular}{cclcccccc}
\hline\hline
Run & c$\star$  & $\beta$  &
$\epsilon$SN &
SFR  &
$\sigma_{gas}$ &
$\sigma_{star}$ &
R$_z$/R$_d$ &
 Hot/Cold Gas \\
${}$ & ${}$ & ${}$ & ${}$ & M$_{\odot}$/yr & km/sec & km/sec &
${}$ & volume ratio \\
\hline
DWFiso01     & 0.05 &  0.05 & 0.1 & 0.20 & 12 & 12 & 0.23 & 0.5 \\
DWFiso02$^a$ & 0.05 &  0.05 & 0.2 & 0.16 & 14 & 14 & 0.31 & 0.6 \\
DWFiso03$^a$ & 0.05 &  0.05 & 0.4 & 0.12 & 17 & 16 & 0.22 & 0.6 \\
DWFiso04$^a$ & 0.05 &  0.05 & 0.6 & 0.06 & 35 & 22 & 0.27 & 0.9 \\
DWFiso05 & 0.20 & 0.05  & 0.2 & 0.18 & 22 & 20 & 0.43 & 1.0 \\
DWFiso06 & 0.20 &  0.10 & 0.2 & 0.15 & 20 & 16 & 0.61 & 1.5 \\
DWFiso07 & 0.20 &  0.20 & 0.2 & 0.13 & 18 & 20 & 0.5  & 2.0 \\
DWFiso08 & 0.20 &  0.20 & 0.6 & 0.03 & 36 & 28 & 0.84 & 1.6 \\

\hline
\end{tabular}
\caption[Summary of the properties of isolated dwarf DWFiso runs]
{Summary of the properties of some of the isolated dwarf runs for a
given set of c${\star}$, $\beta$ and $\epsilon$SN parameters.  SFR is
measured over the whole disk after it has set to a quasi steady
rate. $\sigma_{gas}$ and $\sigma_{star}$ (the azimuthal velocity
dispersions) and $R_z/R_d$ are measured at two stellar disk scale
lengths. The volume ratio between the hot gas (T $> 10^5$ K) vs colder
gas included all gas particles within 2~kpc from the disk
plane. ${}^{\rm a}$ Parameters used for the Cosmological Runs}
\label{tbl-3}
\end{table*}

\noindent where $M_{SNII}$ is the mass of supernovae produced in a
star in a given timestep, r is the distance from the star to the gas
particle in question and $\rho_{ave}$ is the local gas density and
$\beta$ is a normalization factor. In \S2.3.2, we explore $\beta$
values in the range 0.05 to 0.2.  For each SF event the maximum number
of particles that can have the cooling disabled as from eq.(4) is the
number of SPH neighbors (32 in our simulations).

Once formed, we treat each star particle as a single stellar
population with uniform metallicity. The code also keeps track of mass
loss from stellar winds. With the adopted IMF, star particles lose up
to 30-40\% of their original mass as their underlying stellar
population ages. This mass gets distributed to nearby gas
particles. Metals come from both SNe Ia \& II.  Metal enrichment
follows \citet{raiteri96}.  Like energy, metals are distributed using
the smoothing kernel over the 32 nearest neighbor particles.  We plan
to study the metallicity of the baryonic component of our simulated
galaxies and  to introduce metal lines cooling in future papers.

\subsection{Using Isolated Galaxy Models to Calibrate the SF and SN Feedback Algorithms}

We performed a number of simulations of isolated in equilibrium galaxy
models to study the effect of different combinations of the three free
star formation parameters (c$\star$, $\beta$ and $\epsilon$SN) on the
properties of galaxies with a quiescent star formation and over a
range of masses. Parameters were varied within a range of plausible
values suggested by observational constraints.  The star formation
efficiency parameter c$\star$ was varied over the range [0.01 - 0.4],
the mass factor $\beta$ over [0.05--0.4] and the SN efficiency
$\epsilon$SN in the range [0.1 -- 0.6] (see Table 1). This is a subset
of the parameter space explored in S06 where tests were focused on a
Milky Way (MW) sized galaxy.  However, it is important to extend tests
to less massive galaxies. Hence in this work we followed a
complementary approach, focusing on the properties of a much smaller
galaxy and then verifying that the best parameters so identified would
model realistically star formation in a Milky Way sized halo.

\subsubsection{The Galaxy Models}
To explore the effects of our feedback recipe we applied it to two
galaxy models having different peak circular velocities: 220
(MWiso) and 70 km/sec (DWFiso) respectively (where iso stands for
``isolated''). Both models were built as equilibrium configurations
using the procedure outlined in \citet{springel00}.  Rather than
trying to model specific galaxies, we built models consistent with the
trend for smaller galaxies to have smaller bulge components and a
larger cold gas fraction in the disk \citep{mcgaugh05,west05}. The
models include a rotationally supported stellar and gaseous disk, a
bulge for the MW model only and a dark matter halo component extending
to the virial radius.  Stars start with a Toomre parameter Q=2. Disks
are built to be stable to bar instabilities.
  
The dwarf galaxy has no bulge component and gas contributes 50\% to
its total disk mass.  10$^5$ particles are in its dark matter halo,
and 1.5 $\times$ 10$^4$ gas and 2 $\times$ 10$^4$ star particles are
in its disk.  The softening length is set to about 0.2 times the disk
scale length (R$_d$ $\sim$ 1~kpc). Gas and stars have the same
exponential radial distribution. Due to its relatively shallow potential, we
expect this galaxy to be fairly sensitive to the details of the
feedback and star formation algorithm.

Our Milky Way model (MWiso, the same as in S06) has the structural
parameters outlined in \citet{klypin02}, so providing an excellent fit
to the existing data of the matter distribution in our own Galaxy. Our
Milky Way disk is modeled by 4.5 $\times$ 10$^4$ star particles and 10$^4$ gas
particles with a softening of 325~pc.  This resolution is similar to
that of our cosmological runs.
The stellar bulge to disk ratio is 1:7, and gas contributes about 10\%
of the total mass in the disk component.  Disk scale lengths are 3.5
and 7 kpc respectively for stars and gas \citep{broeils97}.

\subsubsection{Comparison Between Models and Observations}

A set of 25 runs allowed us to study in more detail the effect of
the three free parameters (c$\star$,~$\beta$ and $\epsilon$SN) in our
star formation algorithm on the properties of the DWFiso model. In
this sub-section, we highlight the main trends of simulated galaxy
properties with each of the three parameters to motivate our best
choice parameters.

Table 1 summarizes the results of 8 representative DWFiso runs.  The
parameters used in these 8 runs all produce galaxies with star
formation rates (SFRs) that range between 0.03 and 0.20 M$_\odot$/yr
and have velocity dispersions of their cold (T $<$ 30.000 K) gas
component that range between 12 and 36 km/sec. These values are all
consistent with those observed for nearby dwarf galaxies, aside from
the 2 galaxies ($\epsilon$SN = 0.6) with the largest
$\sigma_{gas}$. The SFR measured from the SDSS \citep{brinchmann04}
for small galaxies with stellar masses around 10$^9$ M$_\odot$ ranges
typically between 0.02-0.5 M$_{\odot}$/yr and has a median value of
0.2-0.3 M$_\odot$/yr \citep{brinchmann04}. Observed values for the
velocity dispersion of the cold gas component in disks range in the
10--30 km/sec range \citep{pizzella04}.

\begin{figure}
\begin{center}
\resizebox{8cm}{!}{\includegraphics{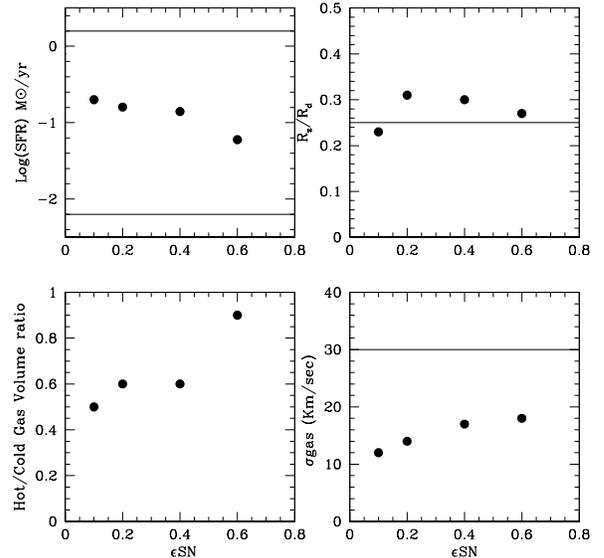}}\\
\end{center}
\caption{Properties of the dwarf galaxy model as a function of
$\epsilon$SN.  The efficiency parameter c$\star$ and the mass factor
$\beta$ have both been set to 0.05. Upper left: Star formation
rates. The two horizontal lines shows the 95\% contour for galaxies of
stellar mass 10$^{10}$ M$\odot$ from the SDSS
\citep{brinchmann04}. Upper right: Disk scale length R$_d$ vs. the
disk scale height R$_z$. The horizontal line is the typical value from
\citet{bizyaev02} and \citet{yoachim06}.  Lower Left: The hot/cold gas
volume ratio as a function of $\epsilon$SN. Lower right: Cold gas
velocity dispersion. The horizontal value is the upper value of
$\sigma_{gas}$ from \citet{pizzella04}.  }
\label{fig:zdisk}
\end{figure}

Although all of the representative runs produce reasonable SFRs and
$\sigma_{gas}$, only runs with low c${\star}$ (0.05 in the
representative sample) produce disks that are thin enough to match
observations.  The observed ratio of the stellar disk scale height
(R$_z$) to stellar disk scale length (R$_d$) is usually in the range
0.25 - 0.33 for normal disk galaxies \citep{bizyaev02}. We find that
values of c${\star}$ larger than 0.05 tend to form ``thick'' disks
with a R$_z$/R$_d$ ratio larger than observed, possibly because more
stars are allowed to form in regions further away from the disk plane.
The R$_z$/R$_d$ ratio is only weakly dependent on $\epsilon$SN or
$\beta$ (runs 1 through 4, 5,~6 \& 7 in Table 1).  This suggests that
at our resolution low values of c$\star$ should be preferred to obtain
thinner disks. Tests in S06 show c${\star}$ to be weakly
dependent on resolution when at least a few tens of thousands gas +
star particles are used. However, higher c${\star}$ values have been
used when star forming regions are individually resolved
\citep{tasker05}.

As the $\beta$ factor is increased, more gas is affected by feedback
and the volume ratio of the hot/cold phases increases (runs 6,7,8 in
Table 1).  In our tests we found that runs with the required low
values of c$\star$ tend to form and maintain a large bubble of hot gas
in the central region of our models, locally inhibiting star
formation, unless $\beta$ is also low. We thus excluded the region of
parameter space with $\beta >$ c$\star$ from our analysis as being
unrealistic.

Figure 2 shows how the gas spatial distribution of the DWFiso model is
affected as c${\star}$ and $\beta$ are fixed to their desired value of
0.05 and $\epsilon$SN is varied from 0.1 to 0.6. In the DWFiso runs,
higher $\epsilon$SN leads to lower star formation rates and a higher
cold gas turbulence (e.g runs 3, 4, and 8). The SFR of the dwarf galaxy
model is most sensitive to the SN energy efficiency parameter
$\epsilon$SN, with a realistic SFR obtained with values in the range
0.1-0.6. The images in Figure 2 show that large SN efficiency values
create a ``galactic fountain'' and a patchy gas distribution on the
disk plane resembling that of observed small galaxies. However, even
with $\epsilon$ = 0.6 only a very small fraction of the gas ejected
from the disk plane becomes unbound from the dark matter halo, in
agreement with observational evidence and some theoretical
expectations \citep{maclow99,martin99,mayer04b,mcgaugh05}.

\begin{figure}
\begin{center}
\resizebox{8.5cm}{!}{\includegraphics{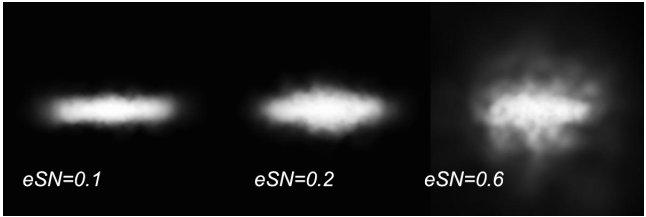}}\\
\end{center}
\caption{ The gas projected density for the dwarf galaxy model seen
edge on as $\epsilon$SN is varied. The box is 20kpc across. }
\label{fig:feedback}
\end{figure}

\begin{table*}
\centering
\begin{tabular}{cclcccccc}
\hline\hline
Run & c*  & $\beta$  &
$\epsilon$SN &
SFR  &
$\sigma_{gas}$ &
$\sigma_{star}$ &
R$_z$/R$_d$ &
 Hot/Cold Gas Volume Ratio \\
${}$ & ${}$ & ${}$ & ${}$ & M$\odot$/yr & km/sec & km/sec &
${}$ & ${}$ \\
\hline
MWiso1       & 0.05 & 0.05 & 0.1 & 0.85 & 12 & 13 & 0.04 & 0.22 \\ 
MWiso2      & 0.05 & 0.05  & 0.4 & 0.8 &  18 & 20 & 0.05 & 0.26 \\ 
MWiso3     & 0.05 & 0.05 & 0.6 & 0.7  & 20 & 24 & 0.05   & 0.4 \\
\hline
\end{tabular}
\caption[Summary of the properties of isolated Milky Way runs]
{Summary of the properties of the isolated Milky Way runs: quantities defined as in Table 1}
\label{tbl-3}
\end{table*}

Tests on the MWiso model in S06 preferred values of c$\star$ in the
0.01--0.1 range but did not constrain $\epsilon$SN significantly due
to the relatively deep potential.  For the MWiso model we set both
c$\star$ and $\beta$ to 0.05 and varied $\epsilon$SN in the range 0.1
to 0.6 (Table 2). As in S06 the effect of injecting more energy into
the ISM is much smaller than for DWFiso, due to the deeper potential
well. Increasing $\epsilon$SN by a factor of 6 has a small effect on
the properties of the ISM, with $\sigma_{gas}$, and the hot/cold gas
volume ratio roughly doubling. A smaller ``galactic fountain'' is
created, but only a small fraction of the gas is ejected away from the
disk and none actually leaves the boundaries of the surrounding dark
matter halo.  The vertical velocity dispersion of the stellar disk
increases slightly while the SFR decreases by less than 20\% across
the whole disk. Disk galaxies over a range of stellar masses show a
remarkably tight relation between the gas surface density
$\Sigma_{gas}$ and the local surface density of star formation,
$\Sigma_{SFR}$.  Equation 4 of \citet{kennicutt98} (originally
  formulated for the average properties of individual galaxies)
states the exact form of this relationship:

\begin{equation}
\Sigma_{SFR} = (2.5 \pm 0.7) \times 10^{-4} (\frac{\Sigma_{gas}}{1 M_{\odot} pc^{-2}})^{1.4\pm0.15} M_{\odot} yr^{-1} kpc^{-2}
\end{equation}

 As expected from the formulation of eq.5 \citep{kravtsov05}, star
formation across the whole disk of our MWiso models follows closely
the slope of the observed  Kennicutt's law and is shown in Figure
3. Increasing $\epsilon$SN lowers the SFR normalization slightly,
again in agreement with the finding from S06.
  
To summarize the set of tests described in this sub-section shows
that our SF + SN feedback scheme provides a reasonable match to some
of the basic properties of present day, isolated galaxies: Kennicutt's
law, the observed SFR of galaxies, the turbulence of cold gas (at
scales of a few kpc) and the thickness ratio of stellar disks. This
provides us with some confidence that our simple scheme is able to
capture the important aspects of SF and its effects on the ISM of
galaxies over a significant range of masses. There is some degree of
interplay between the parameters that could be completely captured
only by an even larger set of tests. However, it is clear that within
the range of parameters explored in S06 and here, $\epsilon$SN
predominantly regulates the SFR and the turbulence of the gas. $\beta$
has to be relatively small to avoid unwanted artificial effects and
c$\star$ plays a relevant role in creating ``thin'' stellar disks.
Given all of the above considerations we have chosen to run our
cosmological simulations keeping $c^{\star}$ and $\beta$ fixed at 0.05
while varying $\epsilon$SN in the range 0.2 -- 0.6.

\begin{figure}
\begin{center}
\resizebox{7cm}{!}{\includegraphics{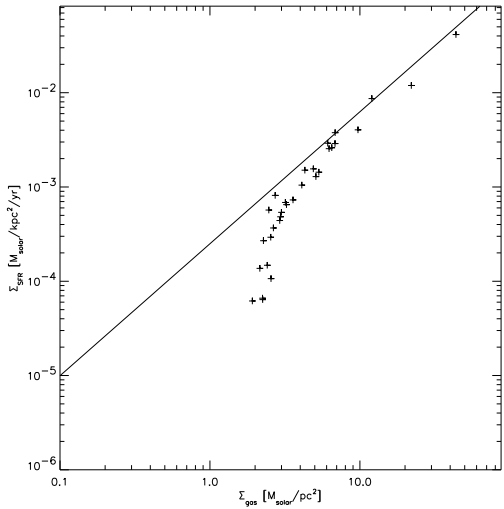}}\\
\end{center}
\caption{ Star Formation rate vs local gas surface density. Dots from the
MWiso3 galaxy model run ($\epsilon$SN=0.6). The straight line is a fit to
Kennicutt's Law. }
\label{fig:MW09}
\end{figure}

\section{Cosmological Runs}

 For our cosmological runs we selected three galaxy-sized halos (DWF1,
 MW1 \& GAL1) from 28.5 and 100 Mpc boxes, low resolution, dark matter
 only simulations run in a concordance, flat, $\Lambda$-dominated
 cosmology: $\Omega_0=0.3$, $\Lambda$=0.7, $h=0.7$, $\sigma_8=0.9$,
 shape parameter $\Gamma=0.21$, and $\Omega_{b}=0.039$ (Perlmutter et
 al. 1997, Efstathiou et al. 2002).  Given the masses and virial radii
 of the selected halos the sizes of the starting boxes are large
 enough to provide realistic torques.  The power spectra to model the
 initial linear density field were calculated using the CMBFAST code
 to generate transfer functions (Zaldarriaga \& Seljak 2000).  These
 three halos were resimulated at higher resolution using the volume
 renormalization technique (Katz \& White 1993). 

We resimulated DWF1, MW1, and GAL1 to have a similar dynamical range
 in each simulation. Each galaxy model was resimulated to have a
 similar number of particles within the virial radius, and the
 softenings were rescaled so that the spatial resolution of each
 galaxy was a similar fraction of its virial radius. With our choices
 of particle number and softening, the smallest subhalos resolved have
 typical circular velocities of 20\% of their host.  For all particle
 species in the high resolution region, the gravitational spline
 softening, $\epsilon(z)$, evolved comovingly from the starting
 redshift (z $\sim$ 100) until z=9, and remained fixed at their final
 value from z=9 to the present. The softening values chosen are a good
 compromise between reducing two body relaxation and ensuring that
 disk scale lengths and the central part of dark matter halos will be
 spatially resolved (see \citealp{diemand04} for a number of relevant
 tests).  Integration parameter values were chosen as suggested in
 Moore et al. (1998) and then confirmed in Power et al. (2002).

\begin{table*}
\centering
\begin{tabular}{cclcccccc}
\hline\hline
Run & Virial Mass  & Virial Radius  &
Vc$^{a}$ & Formation time  & $\lambda$ & 
Last major merger &
 $\epsilon$ &
 N$_{tot}^{b}$ at z=0 \\
${}$ & M$_\odot$ & kpc & km/sec & z &  & z &
kpc & dark+gas+stars \\
\hline
DWF1 &1.6 10$^{11}$  & 142  &   70  &   1.5   & 0.01  & 2.3  &  0.3  & $\sim$ 860.000 \\
MW1 & 1.15 10$^{12}$  & 271  &  134    &   0.6   & 0.07  & 2.5  & 0.6-0.3$^1$  & $\sim$ 700.000 - 4.000.000$^1$ \\
GAL1& 3.1 10$^{12}$ &  380    &  185    &  0.96  & 0.035 & 2.75  & 1.  & $\sim$ 480.000  \\

\hline
\end{tabular}
\caption[Summary of the properties of the three cosmological halos]
{Summary of the properties of the three cosmological halos: $^a$
Circular velocity at virial radius, $^b$ number of gas and star
particles changes slightly depending on $\epsilon$SN. $^{1}$ smaller/larger values are for
the z=0.5 Super High Res.  Run. respectively }
\label{tbl-3}
\end{table*}

  We selected a halo mass range to study halos associated with typical
 disk galaxies. The three selected halos have masses of 1.6 10$^{11}$
 (DWF1), 1.15 10$^{12}$ (MW1), and 3.1 10$^{12}$ M$_{\odot}$ (GAL1)
 measured within their virial radius R$_{vir}$ (the radius enclosing
 an overdensity of 100 times the critical density $\rho_{crit}$).
 GAL1, the largest halo, is more massive than that of our Milky Way,
 as a recent analysis points to a Milky Way halo of about 10$^{12}$
 M$_{\odot}$ \citep{klypin02}. This is the mass of our intermediate size
 halo (MW1). Our least massive halo, DWF1 corresponds to that of a
 typical disk field galaxy.  The three galaxies span a range of
 circular velocities V$_c$ (measured at R$_{vir}$) and defined as
 $\sqrt M(r<R)/R$) between 70 and 185 km/sec. Note that this V$_c$ is
 defined differently than the V$_{rot}$ used later in the paper. Dark
 matter particle masses in the high resolution regions were 7.6
 $\times$ 10$^5$, 6.05 $\times$ 10$^6$ and 2.3 $\times$ 10$^7$
 M$_{\odot}$ and the spline softening was 0.3, 0.6 and 1 kpc for DWF1, MW1
 \& GAL1 runs, respectively. The MW1 halo was also run at higher
 resolution to test resolution effects.  The results of the resolution
 testing are in \S8. GAL1 is the same halo described in
 \citet{governato04}. The main halo properties are summarized in Table 3.

The merging histories and angular momentum of parent dark matter halos
play a major role in defining the final properties of the galaxies
that form inside them \citep{cole00}. It is therefore important to make
sure that our halos have merging histories and spin parameters
somewhat representative of the global population.  The three galaxies
were selected with the only criteria of the redshift of their last
major merger (z$_{lmm}$) $>$ 2 and with no halos of similar or larger
mass within a few virial radii. A major merger is defined here as
having a 3:1 mass ratio. The three halos have formation times (defined
as the main progenitor achieving 50$\%$ of its final mass) in the 0.6
-- 1.5 redshift range.  Their spin parameter, $\lambda$ (defined as 
$J E^{1/2}/GM^{5/2}$) varies from
0.01 for DWF1 to 0.05 for MW1, with the average value for cosmic halos
$\sim$ 0.035 \citep{gardner01}.

z$_{lmm}$ is likely a crucial parameter in defining the properties of
a galaxy disk \citep{steinmetznavarro02}. If mergers efficiently
destroy disks, a low z$_{lmm}$ would leave less time to grow a new
stellar disk from newly accreted gas \citep{baugh96}.  We therefore
stress that the z$_{lmm}$ of these three halos is close to the average
of the population of galaxy sized halos in our adopted cosmology, as
measured in a recent large set of N-body simulations \citep{li05}.
The same authors showed that previous estimates of the average
z$_{lmm}$ estimates based on the extended Press \& Schechter approach
were biased towards lower z's.  DWF1 has its last major merger at
z=2.3, when a head--on encounter generates a very prolate halo. It
only accretes relatively small halos after that. After its last major
merger at z = 2.5, MW1 has a counter rotating minor merger at z $=$
2. A relatively high z$_{lmm}$ for the MW1 halo is consistent with the
Milky Way forming several Gyrs ago (Wyse 2002). At z $=$ 2.75, GAL1
undergoes multiple mergers and accretes several satellites after
that. A few get disrupted after several passages through its disk.  In
the future, we plan to simulate the formation of galaxies with a wider
range of initial conditions.

For each halo a sequence of runs of increasing complexity was
performed. Table 4 summarizes these runs. For the MW halo initial
conditions, we performed the following runs: DM only (MW1dm), no gas
cooling or SF (MW1ad), no feedback and no UV (MW1g0), UV turned on but
without stellar feedback (MW1g1), and three runs with various
supernova efficiencies (MW1g2 to MW1g4).  In all runs with SF we kept
the star formation efficiency fixed and the mass factor $\beta$ and
varied only the fraction of energy from SN dumped into the ISM
$\epsilon$SN from 0.2 to 0.6. Based on the results from \S 2 we will
mainly focus on the properties of the runs with a SN energy efficiency
of 40$\%$ ($\epsilon$SN=0.4).  Results from the cosmological runs
confirm a posteriori this choice: a higher SN efficiency creates a
DWF1 galaxy with an unrealistic low cold gas mass fraction in the disk
($<3\%$).

Our MW1g4  ($\epsilon$SN=0.6) run of the MW1 halo was repeated at
eight times better mass resolution and 2 times better spatial resolution
(run MW1hr in Table 4).  The run was stopped at z=0.5 due to its
computational cost.  With four million resolution elements within the
virial radius, a  star particle mass of 3 $\times$ 10$^4$ M$_{\odot}$ and a
force resolution of 0.3 kpc, this run has likely the highest resolution
ever achieved for a Milky Way sized galaxy simulation carried to a
relatively small redshift. We will use this simulation to test the
effects of increasing resolution in \S 8.

\bigskip

\begin{table}
\begin{tabular}{ccccc}
\hline
run    &  c*    & $\beta$ & UV  & $\epsilon$SN \\
       &        &         &     &             \\ \hline
DWF1g1 & 0.05   & -       & off & off         \\ 
DWF1g2 & 0.05   & 0.05    & on  & 0.4            \\ 
DWF1g3 & 0.05   & 0.05    & on  & 0.6           \\
MW1dm   & dm only &        &                \\
MW1ad   & no SF, no cooling                \\                 
MW1g0  & 0.05   & -       & off & off      \\
MW1g1  & 0.05   & 0.05    & on & off           \\ 
MW1g2  & 0.05   & 0.05    & on  & 0.2             \\
MW1g3  & 0.05   & 0.05    & on  & 0.4                 \\
MW1g4  & 0.05   & 0.05    & on  & 0.6               \\ 
MWhr   & 0.05   & 0.05    & on  & 0.6     \\ \hline
GAL1g1 & 0.05   & 0.05    & on  & off        \\
GAL1g2 & 0.05   & 0.05    & on  & 0.2              \\
GAL1g3 & 0.05   & 0.05    & on  & 0.4        \\ \hline
\end{tabular}
\caption{Main  Parameters of Cosmological Galaxy Runs}
\end{table}

\section{The effect of feedback on the properties of satellites }

\begin{figure}
\begin{center}
\resizebox{8cm}{!}{\includegraphics{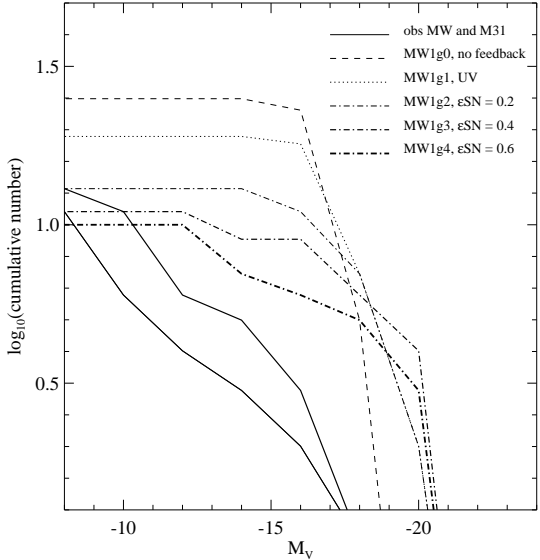}}\\
\end{center}
\caption{The luminosity functions of satellites within R$_{vir}$ in
all 4 MW1 runs, compared with the luminosity functions of both the
Milky Way's and of M31's satellites. Only satellites containing at
least 64 DM particles are included. The 2 runs without supernova
feedback produce far too many satellites.  However, all 3 of the runs
including supernova feedback produce a number of satellites that
reasonably matches that of the Milky Way and M31 satellite populations
(lower and upper solid lines respectively), although the
simulated satellites are still too bright.}
\label{fig:MWLF}
\end{figure}

To explore the effect of UV and supernova feedback on satellites, we
compare the properties of satellites within the virial radius in the
six MW1 simulations described in Table 4. The MW1 satellites are the
most easily comparable to the satellites of our own Milky Way
\citep{mateo98}. Unless specified, similar qualitative conclusions
apply to DWF1 and GAL1.  We will present a much more detailed analysis
of the satellites in a subsequent paper. Subhalos were identified
using SKID \citep{governato97} and only those with at least 64 DM
particles were used in the analysis.  That minimum number of dark
matter particles translates to minimum dark matter masses of: 0.49,
3.9, and 14.9 $\times$ 10$^8$ M$_\odot$ solar masses for the
satellites of DWF1, MW1, and GAL1 respectively.  The resolution limit
translates into a circular velocity limit of V$_c$ $\sim$ 20--30
km/sec depending on the individual satellite density profile.  We use
the ages and metallicities of each star particle in the satellites to
determine their absolute V-band magnitudes assuming no dust reddening.
Satellites have a mean (not luminosity weighted) B-V of 0.63. We
verified that field dwarfs have slightly bluer colors.

Figure 4 shows the satellite luminosity functions of the MW1
cosmological runs compared with those observed for the Milky Way and
M31.  A comparison of both the g0 and g1 simulations (no feedback;
then UV on, but no SN feedback) with the observed satellite luminosity
functions of the Milky Way and M31 highlights the missing satellite
problem.  On the contrary, runs including SN feedback result in
satellite populations similar in number, although not in luminosity,
to the observed populations.  The total number of luminous satellites
is not strongly dependent on the strength of feedback implemented in
the S06 algorithm.

Although both g0 and g1 have a total of 26 satellites, the runs
including supernova feedback all have 20 or fewer satellites total,
only some of which host stars. This difference in total satellite
number occurs because simulations with more severe overcooling form
satellites very centrally concentrated and less susceptible to tidal
effects and complete disruption (see also
\citealt{maccio05}). Similarly, runs with cooling but no SN feedback
(as MW1g0 and MW1g1) contain $\sim$2x the number of satellites as the
DM only run or the run with no gas cooling.

\begin{figure}
\begin{center}
\resizebox{8.25cm}{!}{\includegraphics{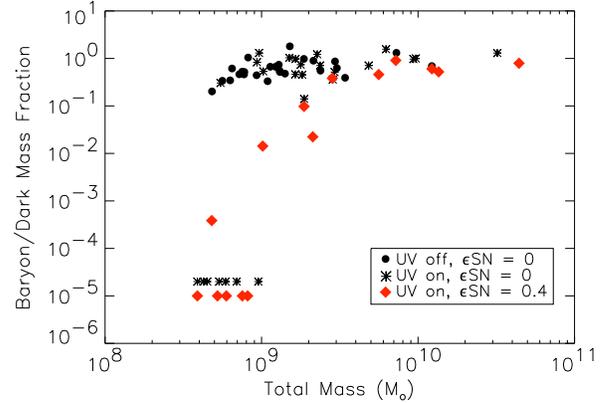}}\\
\end{center}
\caption{The baryon to DM ratio for all the subhalos within R$_{vir}$
of the MW1 galaxy. With no UV field or SN feedback the baryon fraction
is constant down to our resolution limit of 64 DM particles. UV
creates some dark satellites, while SN feedback removes a substantial
fraction of baryons from halos with total masses below a few times
10$^9$ M$\odot$.  Note that total satellites masses have been affected
by tidal stripping. Halos with no baryons in the UV only run have been
shifted upward for clarity.  }
\label{fig:MWLF}
\end{figure}

\begin{figure*}
\centering
\epsfig{file=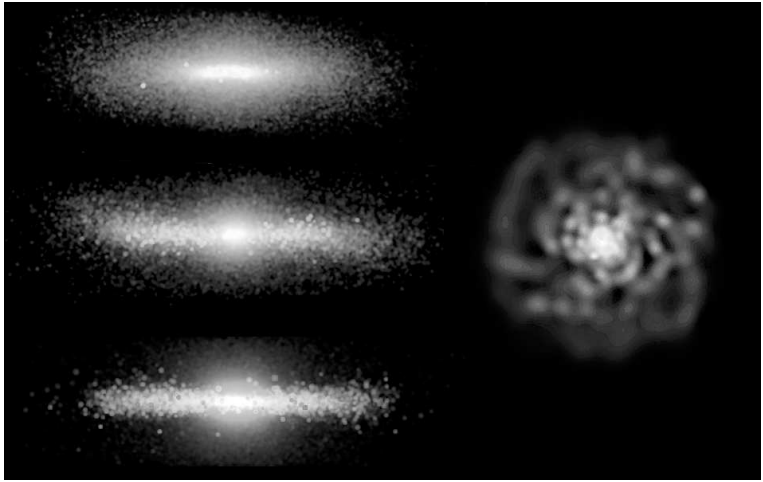,angle=0,totalheight=2.5in,keepaspectratio=true}
\vskip -1mm
\caption{
\label{fig:galaxy}
Left panel: Brightness maps of the edge-on disks of (from top to
bottom) GAL1, MW1 and DWF1 at z=0. Each star particle in the
simulations has been weighted by its age-dependent bolometric
luminosity. Right Panel: the face-on surface density of the gas for
DWF1 at z $=$0. Each frame is 30 kpc across}
\vskip -5mm
\end{figure*}

Figure 5 shows that some of the satellites in the runs including UV
are completely devoid of stars.  There are no satellites within
R$_{vir}$ in any of the simulations that retain gas but don't contain
any stars.  Unlike UV feedback, supernova feedback can also expel gas
from the satellites once they start forming stars, reducing the
overall baryon fraction of satellites containing stars.  These reduced
baryon fraction satellites are seen in Figure 5.  As expected, the
satellites of the run with stronger supernova feedback have a smaller
baryon fraction than the run with weaker supernova feedback.  The
luminosity functions of satellites in the MW1 (and of GAL1)
simulations show that the introduction of SN feedback in particular is
essential to reproduce the observed number of satellites in Milky Way
type galaxies. The number of satellites containing stars appear
is fairly robust as resolution is increased (see Fig.20). At a higher
resolution while tens of less massive dark satellites are resolved,
only a few more are able to make stars. This is because the mass
threshold below which the UV field makes satellites completely dark
\citep{quinn96} is resolved in our runs (Fig.5).  All but one
satellite of galaxy DWF1 (the run with the best mass resolution in our
set) are completely dark when SN feedback is on. Note that the list of
known MW satellites at very low luminosities and  surface brightnesses
might still be incomplete (see \cite{willman04} for a discussion).

\section{Properties of the Disks}

The images in Figure~\ref{fig:galaxy} show that all  three galaxies
have formed a significant stellar disk by redshift z = 0. In this
section, we discuss in detail the spatial and kinematic properties of
the disks, investigate the effects of feedback on disk properties, and
demonstrate the formation of the disks during major, gas-rich mergers.

\subsection{Density Profiles of the Disks}

\begin{figure}
\begin{center}
\resizebox{8cm}{!}{\includegraphics{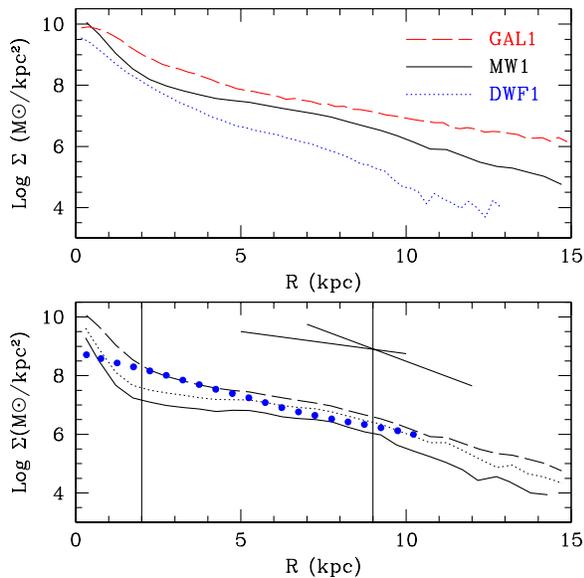}}\\
\end{center}
\caption{The face--on stellar density profile of the three simulated
galaxies. Only stars within 4 R$_z$ from the disk plane are
considered. The upper panel shows the contribution from all stars with
a cosmic age between 0 and 10 Gyrs, roughly corresponding to the K
band \citep{martin05}. The lower panel shows the projected stellar
density profile of the MW1 galaxy for stars in the 0--3 Gyrs (solid),
0--4 Gyrs (dotted) and 0--10 Gyrs (dashed) age range. Dots show the stellar
profile of the MW1hr at z $=$ 1.7. The two straight lines show the
slope of the younger stellar component inside and outside the break
radius.}
\label{fig:profiles4}
\end{figure}

The top panel of Figure~\ref{fig:profiles4} shows the z = 0 stellar
surface density profiles of the disks as a function of radius for our
fiducial runs (DWF1g2, MW1g3, GAL1g2 - $\epsilon$SN=0.4).  Hereafter
we will simply refer to these specific runs as DWF1, MW1, and GAL1 as
we report and discuss our results.  We include all stars within 4 disk
scale heights, R$_{z}$.  The stellar disk component of all galaxies is
well-fit by an exponential distribution between one and three disk
scale lengths. The profiles in Figure~\ref{fig:profiles4} illustrate
that all galaxies also show evidence for a central, steeper component
that extends to 1 or 2 kpc.

 Disk scale lengths, R$_D$, were measured by fitting an exponential
 profile to the disk component, but excluding the steep central
 component and stopping where the surface density shows a break.  To
 approximate the effect of measuring the disk scale length in
 different optical bands, we measured the scale lengths of stars in
 different age ranges: $<$ 2~Gyrs, $<$ 4~Gyrs and $<$ 11~Gyrs, to
 approximate the B, R and K optical bands, respectively
 \citep{martin05}.  These scale lengths are summarized in Table 5.
 The bottom panel of Figure~\ref{fig:profiles4} shows the stellar
 density profiles of MW1 in each of these three stellar age ranges.
 Beyond 2.5--3 scale lengths (shown by the outer vertical line), the
 disk surface density declines rapidly for the youngest stars. These
 disk profiles are quite common in observed spiral galaxies
 \citep{pohlen05,florido01}.  In the remainder of this section we will
 refer mostly to stars younger than 4 Gyr (R band) to illustrate our
 results, unless otherwise specified.

The exponential disk component forms from the inside out as observed
 in many disk galaxies \citep{ryder94}.  However, gas inflow due to
 the late formation of a bar or an oval-like structure in the central
 few kpc, which is responsible for the observed central steepening of
 the profile, can trigger enhanced star formation in the inner disk,
 which then forms partially outside-in. This mode of disk formation is
 significant in run GAL1 and was already reported in
 \citet{governato04}. At z = 0, star formation in GAL1 is primarily
 concentrated in the inner parts of the disk. The scale length of
 stars younger than 2 Gyr in GAL1 is only 1.1 kpc, half that of the
 older component. This is typical of massive S0/early type disk
 galaxies \citep{pohlen04s0}.

\begin{figure}
\begin{center}
\resizebox{10cm}{!}{\includegraphics{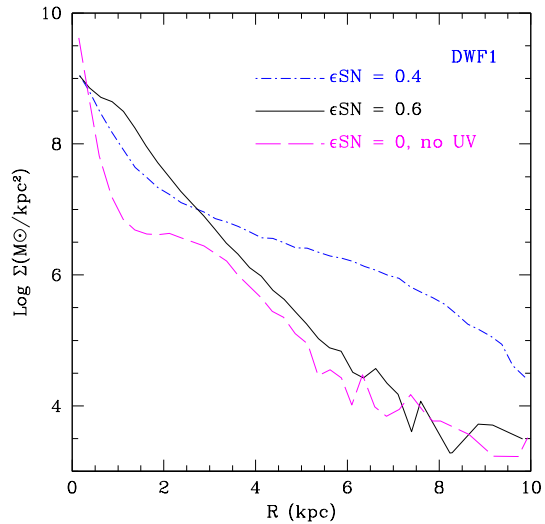}}\\
\end{center}
\caption{The face--on stellar density profile of the DWF1 galaxy as a
function of feedback strength. Only stars within 4 R$_z$ from the disk
plane and in the 0--4 Gyrs age range are plotted. }
\label{fig:profiles5}
\end{figure}

 All three galaxies have extended disks of gas colder than a few
 10$^4$ K with exponential scale lengths larger than even the youngest
 stellar disk component, as in \citet{ryder94}.  The right panel of
 Figure~\ref{fig:galaxy} shows that at our sub-kpc resolution and with
 our feedback scheme we are able to qualitatively capture the complex
 nature of the ISM. The gas distribution in the disk plane of our
 smallest galaxy (DWF1) and in the high resolution MW (MWhr) shows a
 complex structure of spiral arms linked by filaments and surrounded
 by a hotter medium.

The stellar profile of the final disks in the more massive galaxies
(MW1 and GAL1) is only weakly dependent on the strength of supernovae
feedback as we allow the energy efficiency to vary only by a factor of
three, being constrained by the early tests we carried in \S2.
\citet{okamoto05} showed how the density distribution might depend
substantially on the feedback adopted. It is likely that the energy
injection per unit mass in the feedback scheme adopted by Okamoto et
al. is higher than in our simulations.  Instead, the disk stellar
profile of DWF1, our less massive galaxy, is more affected when
varying the amount of SN feedback in the same range (Fig.8). When $\epsilon$
increases to 0.6 the stellar density profile of younger stars in DWF1
is best described by a single, exponential profile with a shorter
scale length and truncated at about 2 R$_d$. Similarly to the
tests performed in \S 2 the cold gas is much more turbulent, making
star formation inefficient at low densities. When feedback is turned
off cold gas settles in a smaller, more compact disk that originates a
denser central core.  As the disk surface density increses the disk
becomes bar unstable redistributing the stellar material and
originating the ``flat'' stellar profile at $\sim$ 2kpc that is
usually associated with strong dynamical instabilities
\citep{debattista05}. In all three galaxies runs without feedback form
``hotter'' stellar disks with a higher scale lenght.

\subsection{Kinematics of the disks}

\begin{figure}
\begin{center}
\resizebox{8cm}{!}{\includegraphics{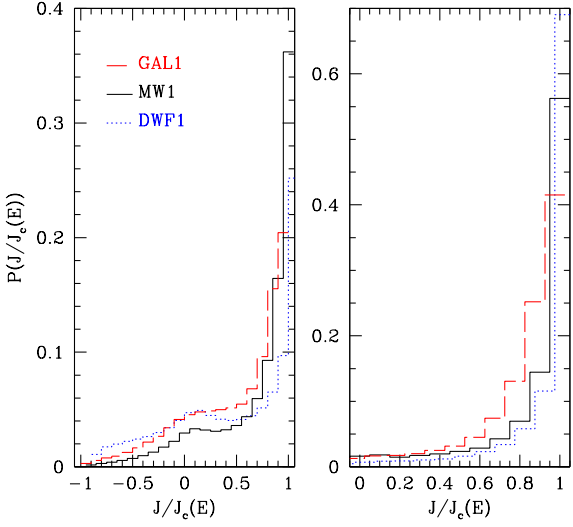}}\\
\end{center}
\caption{Mass weighted probability distributions of the orbital
circularity J/J$_c$ for disk stars in DWF1 (solid), MW1 (dotted) and
GAL1 (dashed). The right panel (note the change of scale) shows only
star particles formed in the last 3 Gyrs.  }
\label{fig:jplot3}
\end{figure}

We first studied the dynamical properties of the simulated objects
and verified that the disks are indeed supported by rotation. For
every star particle within the ``disk'' we computed J$_z$, the
component of angular momentum parallel to the total angular momentum
of all disk stars.  We defined the disk stars as those within 4 (R
band) scale lengths and heights centered on each galaxy. We compared
 J along the z axis to the angular momentum of the co--rotating circular orbit
with similar orbital energy, J$_c$(E). Stars in a disk component
completely supported by circular motions should by definition have
J/J$_c$ = 1.  Figure~\ref{fig:jplot3} shows the mass weighted distribution for all the
disk star particles (left panel) and for disk star particles formed in
the last 3 Gyrs (right panel). The histogram distributions show
clearly that in all three galaxies the disk component is 
mostly supported by rotation.  Our DWF1 and MW1 galaxy models have a
higher fraction of stars on circular orbits when compared with the
distributions showed by \citet{abadi03a} and \citet{okamoto05}. In
particular the young stellar component is heavily dominated by stars
on almost circular orbits: 80\% of young stars in the DWF1 run have
J/J$_c$ $>$ 0.8.  62\% of all stars in the MW1 galaxy have J/J$_c$ $>$
0.8. As a comparison most thin disk stars in the MW have J/J$_c$
$>0.7$ \citep{nordstrom04}. A central hump in the J/J$_c$
distribution, that has often been interpreted as the sign of a massive
halo/bulge component is evident in the DWF1 run. We verified that the
low angular momentum stars are also the oldest, and were formed during
DWF1's last major merger.

The amount of stellar rotation versus velocity dispersion, $v/\sigma$,
is another good measure of how realistic are the disks that we
obtained: V$_{rot}$/$\sigma$ $\sim$ 4 for the MW1 disk stars younger
than 4 Gyrs, where $\sigma$ is the rms average between the tangential and
radial velocity dispersions.  Typical observed values range from 2 to
5 for the disks of normal galaxies \citep{pizzella04}.  Fig.10
shows V$_{rot}$/$\sigma$ for the young stellar component of DWF1.  The
dynamically coldest stellar disk is obtained with $\epsilon$SN
= 0.4. Stronger feedback makes the ISM more turbulent, causing stars to
form with a higher velocity dispersion. For all three galaxies runs
without SN feedback generate stellar disks with lower
V$_{rot}$/$\sigma$ ratios, due to stronger loss of angular momentum
driven by bar instabilities or because the velocity dispersion is
increased by the overall stronger heating by satellites (that survive
disruption much closer to the galaxy center).  These statements hold
true irrespective of the selected age interval for the stars.

\begin{figure}
\begin{center}
\resizebox{10cm}{!}{\includegraphics{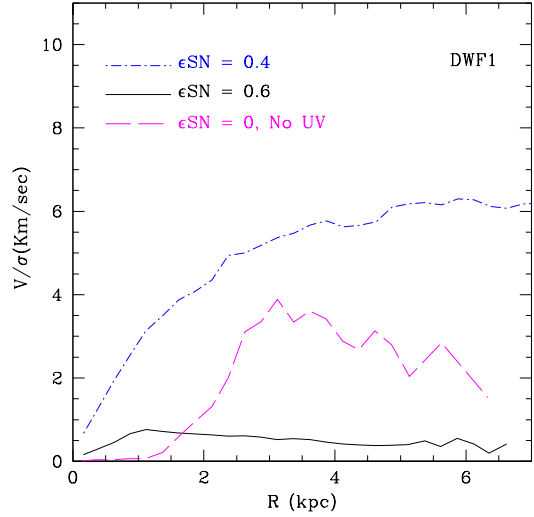}}\\
\end{center}
\caption{The V/$\sigma$ ratio for the young stellar component of DWF1
as feedback is increased: Long dashed: no feedback or UV, dot dashed:
$\epsilon$SN = 0.4, solid: $\epsilon$SN = 0.6}
\label{fig:vsigma}
\end{figure}

\begin{figure}
\begin{center}
\resizebox{9cm}{!}{\includegraphics{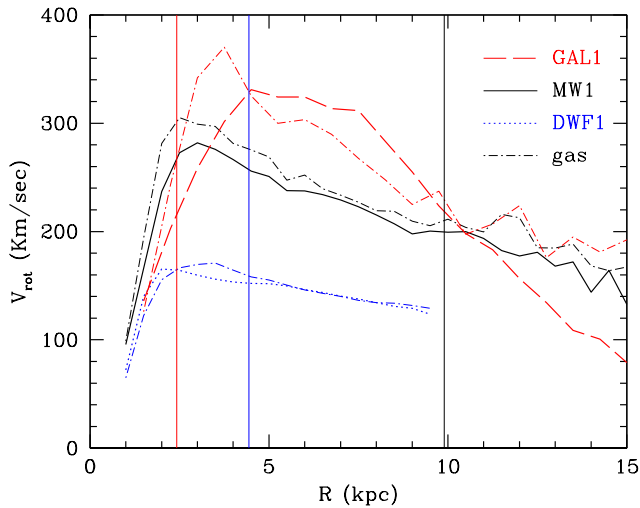}}\\
\end{center}
\caption{Azimuthally averaged rotation velocity of cold gas (dot --
dashed line) and stars ( age $<4$ Gyrs).  From top to bottom:
GAL1,MW1,DWF1. The vertical lines show 2.2~R$_d$ for each galaxy.  }
\label{fig:vrot}
\end{figure}

The azimuthally averaged rotation curves of galaxies are shown in
  Figure~\ref{fig:vrot}.  Both the rotation curve of the cold gas
  component and of the disk stars younger than 4 Gyrs are plotted. For
  both the MW1 and DWF1 runs intermediate age stars and cold gas have
  very similar profiles, confirming that younger stars are in orbits
  close to circular.  In the GAL1 run (our most massive galaxy), the
  stellar and gaseous disks are not completely coplanar. This offset
  creates a difference in the central part of the velocity profile of
  the two components. The outer part of GAL1's stellar disk is also
  dynamically hot, which makes V$_{rot}$ of the stellar component
  decline very rapidly at outer radii.

 While all our galaxies show long lasting bar instabilities, the
 introduction of SN feedback increases the stability of the disk and
 thus reduces the non-axisymmetry of the stellar disk. Stronger
 feedback generates stronger disk axisymmetry. In all simulations with
 feedback, only an oval distortion with scale smaller than the disk
 scale length is present as opposed to a clear bar with scale length
 $\sim$ disk scale length in models without SN feedback. The small
 oval distortion is likely an unresolved bar, because our high
 resolution control run (MW1hr) does display a resolved bar at its
 center.  Disks with feedback are more stable to bar formation because
 their stellar disk builds up more slowly over time. 

 We can understand the stabilizing effect of feedback by comparing the
 mass in the halo to that in the disk using the parameter
 $\varepsilon$= (V$_{max}$ $\times$ R$_D$) / GM$_d$, introduced by
 \citet{efstathiou82} to measure the susceptibility of disks to bar
 formation.  $V_{max}$ is the peak circular velocity and $M_d$ and
 $R_D$ are, respectively, the mass and exponential scale length of the
 disk. Numerical experiments \citep{efstathiou82, mayer04a} show that
 $\varepsilon > \alpha$, with $0.95 < \alpha < 1.1$, is required for
 stability against bar formation of a disk of gas and stars (the
 higher  the gas fraction, the higher $\alpha$).  Measuring
 $\varepsilon$ in a cylinder within 0.5 kpc of the disk plane, we find
 that in the runs without feedback  $\varepsilon$ is already $< 0.9$
 at $z > 1$. In the runs with feedback, $\varepsilon$ does not approach unity
 until near  $z=0$. In runs without feedback the disk is more massive
 and denser at earlier epochs, and thus more bar-unstable. Without
 feedback, all the gas that is accreted rapidly makes disk stars,
 while with feedback a significant fraction of the baryons remain hot
 and diffuse and do not form stars even when they reach the disk.
A visual comparison with real rotation curves shows that our
 simulated galaxies produce rotation curves that raise rapidly and
 then decline too fast at R $>$ 4 R$_d$.  Only DWF1 has a rotation
 curve close to being ``flat'' as observed for most normal spiral
 galaxies \citep{catinella06}. This is likely due to a combination of
 a still too massive central component and the outer part of disks being
 affected by two body heating from halo particles.

\begin{table*}
\centering
\begin{tabular}{ccccccccccccc}
\hline\hline
Run &  Rd$_B$   & Rd$_R$  & Rd$_K$ & Rz$_R$ & Rd$_{gas}$ & Rz$_{gas}$ &   B$_{Tot}$ &I$_{tot}$ &  K$_{Tot}$ & SFR$^1$ M$\odot$/yr & V$_c$$^2$ & M$\star$\\
\hline
DWF1  & 2.02   & 1.95  & 1.2   &  0.3   & 7.  & 0.4  & -19.8 & -21.2 & -22.6 & 0.6 & 160 & 1.38 $\times$ 10$^{10}$\\
MW1   & 4.5    & 3.8   & 2.1   &  1.    & 10  & 1.   & -21.1 & -22.6 & -24   & 2.2 & 205 & 4.6  $\times$ 10$^{10}$\\
GAL1  & 1.1    & 1.2   & 2.15  &  2.3   & 6.  & 1.3  & -22.0 & -23.5 & -24.9 & 7.1 & 180 & 1.36 $\times$ 10$^{11}$ \\
\hline
\end{tabular}
\caption[Summary of Galaxy properties] {Summary of the properties of
the disks of the galaxies formed in the cosmological runs (only runs with
$\epsilon$SN = 0.4 are shown). $^1$ at z=0, $^2$ at 2.2~R$_d$.}
\label{tbl-3}
\end{table*}

\subsection{Gas Rich Mergers and the Formation of Disks at High Redshift}

With our relatively high resolution we are able to follow the
formation of disks early on in our simulations.  Perhaps surprisingly,
the disk of the MW1 galaxy starts forming immediately after the last
major merger event confirming early results by
\citep{springelhernquist05} and \citep{elmegreen05} that gas rich mergers with
substantial feedback are conducive to the observed early formation of
disks \citep{trujillo05}. In our simulations feedback stops a
substantial fraction of gas from turning into stars during the major
merger. 40\% of the mass of the disk of MW1 is  gas at z $=2$.  With
lots of orbital angular momentum available from the merger, gas cools
and rapidly settles into an extended exponential disk, especially in runs
with stronger feedback and higher resolution. Even for DWF1 and GAL1,
substantial stellar disks form immediately after z$_{lmm}$. Without
feedback a much more compact stellar component forms and becomes
rapidly non-axisymmetric, as described in \S5.2. At z$>$1 stellar
disks are 20-30\% larger when feedback is on, although by $z=0$ their
sizes are virtually identical. V/$\sigma$ measured in the disk of the
high resolution MW1hr run is $\sim 1.6$ at z = 1.5 for stars that form
shortly after z$_{lmm}$. This stellar component is heated by a
combination of the rapidly changing central potential and by a couple of
minor satellite accretions onto the disk that occur before z $=$1.5.
This early disk could be associated with the thick disk components of
present day spiral galaxies \citep{seth05,yoachim05,brook04}.

\subsection{Bulge \&   Disk  Decomposition}

To separate the bulge, disk and halo components we assigned to the
``bulge'' all stars in the age range that would give a
V$_{rot}$/$\sigma$ $<$ 1 and within four disk vertical scale heights
from the disk plane.  This selects an old, slowly rotating stellar
component.  Disk stars were those within four disk scale heights and
scale lengths that were not assigned to the bulge.  Stars not
belonging to the bulge, disk or individual satellites were assigned to
the halo. The kinematically defined B/D mass ratio of the fiducial MW1
is = 0.3 or smaller.  This result does not change for other reasonable
decomposition techniques \citep{governato04}. We only used dynamical
definitions of the galaxy components. A photometric decomposition was
not attempted but given that stars in the disks are younger and then
brighter it is likely that a luminosity weighted estimate would give a
lower B/D ratio.  We speculate that the low B/D ratio of the MW1 is
due to the high angular momentum acquired by cold gas during the last
major merger event.  The DWF1 and GAL1 simulations both formed a more
massive ``bulge''.  This result is likely a consequence of different
physical processes: (i) The DWF1 galaxy has very low angular momentum
(the last major merger at z $=$ 2.3 is almost head--on); (ii) GAL1
forms stars in its central region until the present time, as gas keeps
cooling from the surrounding halo.  Its bulge has fairly blue colors
and is relatively more massive. The relevant scale lengths are $<$ 1
kpc, so we decided to postpone a more detailed analysis of the bulge
components of our simulated galaxies when higher resolution
simulations will be available.

\section{Matching the Tully -- Fisher relation}

The Tully--Fisher (TF) relation links the characteristic rotation
velocity of a galaxy with its total absolute magnitude
\citep{giovanelli97a}. Similarly, the ``baryonic TF'' relation \citep{
  mcgaugh00, mcgaugh05} links characteristic rotation velocity with
total disk mass to account for the fact that less massive galaxies are
more gas rich and thus that their stars only account for a small
fraction of their disk total mass. Both relations imply that the total
disk mass and its radial distribution are closely connected with the
total mass of a halo.  Early simulations matched the slope of the TF
relation, but reported \citep{eke01} difficulties in matching its
normalization. The relatively weak feedback adopted in that work might
have caused galaxies to cease star formation relatively early, leaving
the disks too red and then causing the TF relation
offset. Alternatively, the central rotational velocities of the
simulated galaxies were too high due to an excess of matter at the
center of galaxies possibly due to excessive baryonic cooling and the
subsequent adiabatic contraction of the dark matter component. Even
high resolution simulations like \citet{governato04} and
\citet{abadi03a} have shown a consistent offset from the TF relation
of late type spirals, with galaxies being too centrally concentrated
compared to real ones.  Compared to some previous works our
simulations have sufficient resolution to allow us a much direct
comparison with observational data as V$_{rot}$ is now measured at
only a few kpc from the center. This poses a much stronger constraint
on simulations as they have to reproduce the stellar mass distribution
and kinematics at relatively small scales.

When feedback is included, simulations successfully match both the TF
and baryonic TF relations, this result being possibly the most
important conclusion of the work presented here: it shows that our
simulations produce a realistic distribution of stars, baryons and DM
within the central few kpc. To compare our simulations with the
observed TF relations, we used the rotation curves plotted in Figure 9
(and described in \S5.2) to determine V$_{rot}$ measured at 3.5 disk
scale lengths for DWF1, MW1, and GAL1 as in the Giovanelli sample. We
then obtained the global magnitudes for each galaxy model by coupling
the star formation histories (SFHs) of our simulations with GRASIL
\citep{silva98}, a code to compute the spectral evolution of stellar
systems taking into account dust reprocessing. We considered each star
particle as a single stellar population with the age, metallicity, and
mass assigned to it by the simulation.  Stars with metallicities less
than Z = .0005 were assigned that minimum value. GRASIL models the
distribution of molecular clouds and dust based on the structural
parameters of a galaxy disk and bulge. The quoted magnitudes in Table
5 are the average of different viewing angles.

\begin{figure}
\begin{center}
\resizebox{7.75cm}{!}{\includegraphics{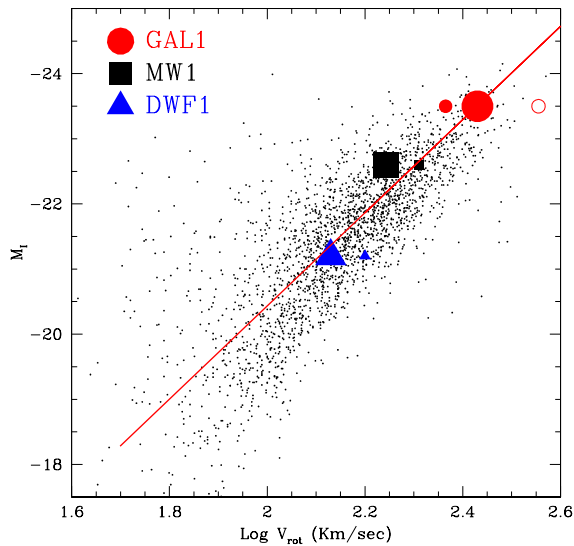}}\\
\end{center}
\caption{ The Tully Fisher relation using the data compilation from
Giovanelli 2005 (private communication) and a fit to
\citet{giovanelli97a}. Solid triangle: DWF1, Solid Square: MW1, Solid
Dot: GAL1. Bigger dots shows V$_{rot}$ measured at 3.5R$_d$. Smaller
dots shows the effect of measuring V$_{rot}$ at 2.2R$_d$. The small open dot uses V$_{rot}$ 
measured from GAL1 cold gas component}
\label{fig:TF}
\end{figure}

\begin{figure}
\begin{center}
\resizebox{8cm}{!}{\includegraphics{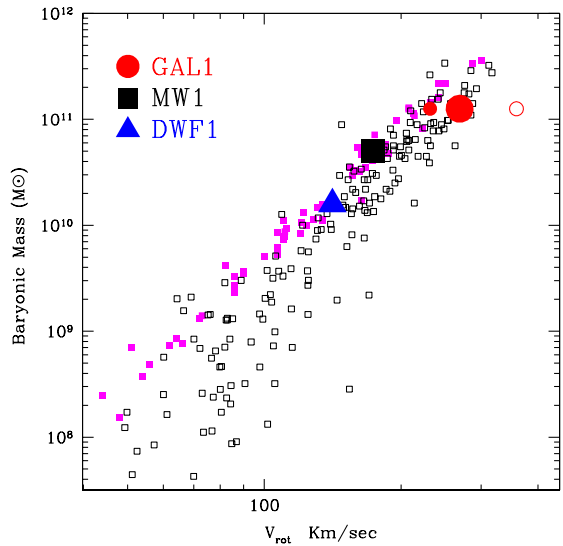}}\\
\end{center}
\caption{ The Baryonic Tully Fisher relation.  Squares show data
points from \citep{mcgaugh05}. V$_{rot}$ is measured at 3.5R$_d$ (as
in the observational sample) for each of the three galaxies.  The two
smaller points for the GAL1 run show the effect of measuring V$_{rot}$
at 2.2~R$_d$ using stars and the cold gas component, as shown in
Figure~\ref{fig:TF}.}
\label{fig:barTF}
\end{figure}

Figures~\ref{fig:TF} and ~\ref{fig:barTF} show our galaxy models
 overplotted on TF and baryonic TF data from Giovanelli (2005, private
 communication) and \citet{mcgaugh05}, respectively.  The opposite
 residuals for DWF1 and MW1 from the average TF relation (continuous
 line in Fig.11) correlate well with their halo spins: lower than
 average for DWF1 and higher for MW1.  As GAL1 dynamical tracers are
 slightly decoupled (see \S 5.2), its location on the TF
 relations somewhat depends on the tracers chosen (old or young stars
 for the disk scale length, stars or cold gas as tracer of the
 velocity field). As GAL1's stellar mass is larger than 10$^{11}$
 M$\odot$ it is a likely candidate for an early spiral or even an S0.
 We speculate that another type of feedback, possibly from an active
 QSO at the bulge center \citep{dimatteo05,hopkins05}, would decrease
 star formation in the center and lower the central gas density,
 reducing the relative weight of the central baryonic concentration.

Improvements in matching the Tully Fisher and in conserving the
angular momentum content of disks have been consistently reported in
the literature as the robustness of simulations and the detail of
physical modelling has continued to improve \citep{sommer03,
abadi03a,governato04,robertson04}.  As we show in \S 8 our simulations
have a softening of the order of only 20\% of the disk scale lengths
and give robust and likely converging results at scales corresponding
to the typical scale length of stellar disks. We verified that our
match with the TF relations is not due to V$_{rot}$ $<<$ V$_c$,
i.e. the disks are not dynamically ``hot'' within a few disk scale
lengths and gas and young stars trace very closely the underlying
potential.  We verified that runs with no feedback produce
galaxies slightly redder, but only in the B band and only for DWF1
where the effect is 0.3 mag with $\epsilon$SN = 0.4 and 0.5 mag for
$\epsilon$SN = 0.6.  For more massive galaxies and/or redder bands the
effect is only 0.1 - 0.05 mag as the effect of feedback is smaller and
star formation happen mostly at higher redshifts. This small color
shift is not sufficient to explain the shift from the I band TF
measured in some previous simulations.

\begin{figure}
\begin{center}
\resizebox{7cm}{!}{\includegraphics{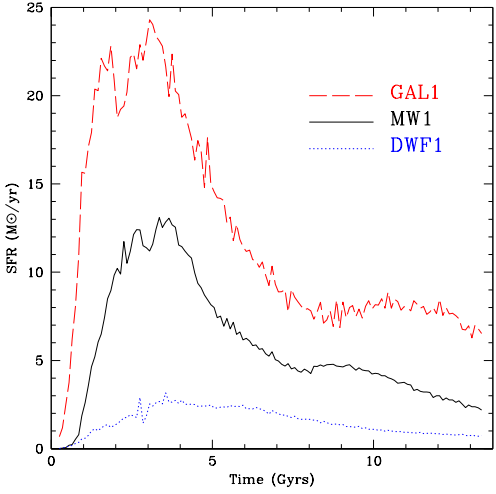}}\\
\end{center}
\caption{ Star formation histories of DWF1, MW1, and GAL1, including
all stars within 4 disk radial and vertical scale
heights. $\epsilon$SN=0.4 and UV field on for all runs.  Solid black:
MW1, long dashed: GAL1, dotted: DWF1. }
\label{fig:sfr}
\end{figure}

\section{Feedback and the Star Formation Histories of galaxies}

Not only does feedback affect the structure and stability of disks, as
discussed in \S5.1 and \S5.2 respectively, but we also expect it to
shape their star formation histories.  The isolated test simulations
of \S2 hinted at this effect, and in this Section we explore the
effect of feedback on the star formation histories of DWF1, MW1, and
GAL1.  The SFHs of bulge and disk stars in these three runs are shown
in Figure 14. The plot shows that the peak of star formation is
delayed in less massive halos, as expected if feedback has a more
pronounced effect in galaxies with shallower potential wells
\citep{maller02}. To understand this result, consider the natural
property of CDM models that star formation at high redshift happens in
a number of progenitors that eventually coalesce to form a present day
galaxy.  We used the DWF1 simulation to verify that increasing
feedback limits the efficiency of star formation in progenitor halos
with V$_c$ $<$ 30-50 km/sec and delays it until small subhalos merge
into halos with a sufficiently deep potential well (see also \cite{neinstein06}).

\begin{figure}
\begin{center}
\resizebox{7cm}{!}{\includegraphics{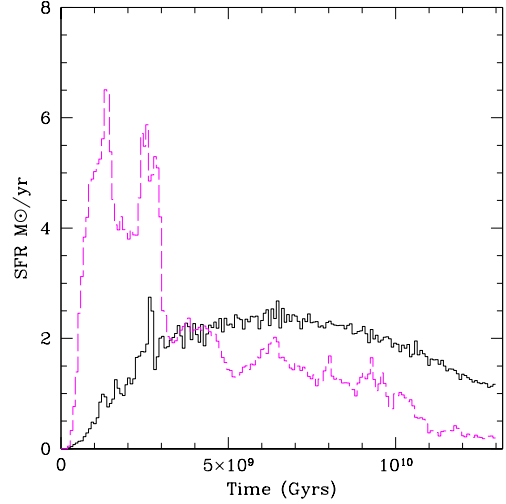}}\\
\end{center}
\caption{Galaxy DWF1: SFH including  all stars within 4 R$_z$ and R$_d$
 from the disk plane for two runs. Solid line: $\epsilon$SN=0.6, long
dashed: no feedback, no UV. The addition of feedback smooths out the
SF peaks  otherwise present at high redshift and during the last
major merger event at z=2.3. Feedback delays the conversion of gas
into stars until gas accumulates and cools in the potential well of
the main progenitor.  }
\label{fig:dwf1sfr}
\end{figure}

To illustrate the crucial effect of feedback on the SFH of DWF1,
Figure~\ref{fig:dwf1sfr} compares its SFH in the absence of feedback
and with $\epsilon$SN = 0.6, the strongest feedback adopted in our
study.  Early star formation is significantly reduced in the case of
feedback, and we no longer see the initial strong starbursts evident
in the run with no feedback. Lowering the efficiency of star formation
during mergers preserves a large quantity of cold gas that rapidly
settled in rotationally supported disks.  The run with feedback has a
present day SFR almost ten times higher than that without.  The SFR in
the run with feedback is far too low to match observations.

\begin{figure}
\begin{center}
\resizebox{8cm}{!}{\includegraphics{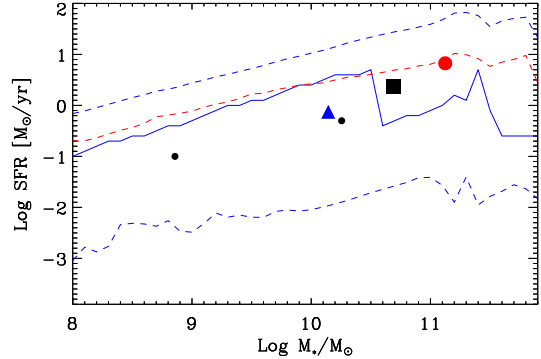}}\\
\end{center}
\caption{Present day star formation rate vs galaxy stellar mass for
both simulated and observed galaxies. Triangle: DWF1, Square:
MW1. Solid Circle: GAL1. The two smaller dots are field galaxies in the
high resolution region of our simulations. The continuous line is the
average SFR for the SDSS sample described in \citet{brinchmann04}. The
central dashed line is the median value. The upper and lower dashed
lines include data three sigma from the mean of the observational
sample}
\label{fig:SDSS}
\end{figure}

The z = 0 star formation rates of DWF1, as well as of the other two
galaxies, are quite similar to those observed in galaxies with the
same stellar mass. Figure~\ref{fig:SDSS} shows the z = 0 star
formation rates of our simulated galaxies overplotted with those
measured for SDSS galaxies in a wide mass range
\citep{brinchmann04}. Two low mass field galaxies outside the virial
radius of galaxy MW1 (but within the high resolution region) have also
been included to show a fair agreement between our theoretical model
and observational data over more than two orders of magnitude in
stellar mass.

\begin{figure}
\begin{center}
\resizebox{7cm}{!}{\includegraphics{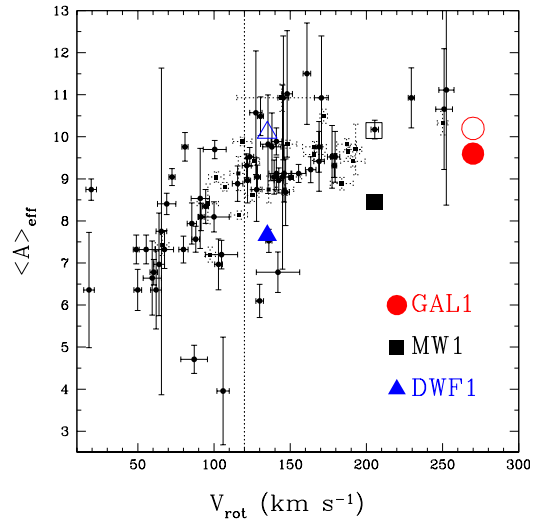}}\\
\end{center}
\caption{Mass weighted average age of stellar
populations vs galaxy stellar mass.  Small dots with error bars: from
\citet{macarthur04}. Solid large dots: simulations with $\epsilon$SN =
0.4. Open dots: simulations with no feedback or UV. V$_{rot}$ is
measured at 3.5 disk scale lengths.}
\label{fig:lauren}
\end{figure}

Our galaxy models run with SN feedback also reproduce the observed
``downsizing'' of field galaxies \citep{cowie96,macarthur04}: galaxies
with a less massive stellar component are also younger.
Figure~\ref{fig:lauren} shows that the inclusion of feedback and its
differential effects at different galaxy masses naturally reproduces
the observed trend: more massive galaxies also have older stellar
populations. This trend disappears when runs with no feedback are
considered (open circles) and average stellar ages correspond to the
time of collapse of the first protogalaxies. As SN feedback removes
baryons and delays star formation in smaller halos, it has a stronger
effect in less massive galaxies that formed from even less massive
progenitors at high redshift. Gas must collect inside a sufficiently
deep potential well before it is able to cool efficiently. More
massive galaxies are formed from the merging of more massive
progenitors, that were able to form stars already at high redshift.
While CDM forms structure from the bottom up, the introduction of
feedback reverses the trend for stellar populations.

Recently, \citet{bell01} and \citet{conselice05} used observations of
spiral galaxies to measure the stellar mass associated with galaxies
of different rotational velocities and halo mass. A comparison of our
simulations with data in \citet{conselice05} and the best fit to
\citet{bell01} is plotted in Fig.18 and shows that our models form the
right amount of stars in their bulge and disk components. We also
verified that MW1 contains 4 $\times$ 10$^{10}$ M$\odot$ of dark matter within
8.5 kpc, again in good agreement with estimates for the Milky Way
\citep{eke01}.  These results  confirm that our modelling of
SF and feedback created galaxies with the right amount of stars for
their halo mass.

\begin{figure}
\begin{center}
\resizebox{8cm}{!}{\includegraphics{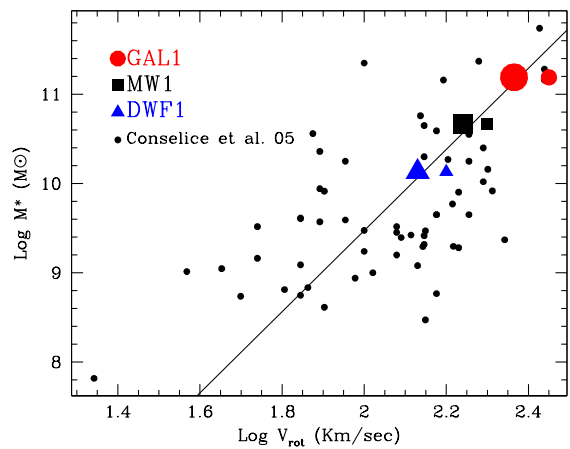}}\\
\end{center}
\caption{Total stellar mass vs. rotational velocity. Solid small dots are
data from \citet{conselice05}. Large dots: simulations measured at
3.5 R$_d$. Small points: simulations measured at 2.2 R$_d$.  The
straight line is a fit from \citet{bell01}.  }
\label{fig:conselice}
\end{figure}

\section{Resolution Tests:  The central mass profile of Galaxies and the luminosity function of satellites. }

\begin{figure}
\begin{center}
\resizebox{8cm}{!}{\includegraphics{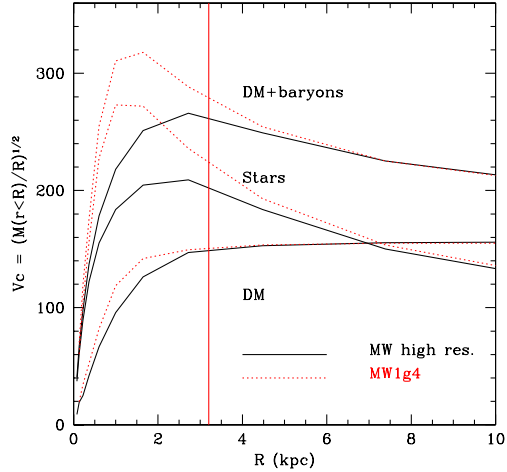}}
\end{center}
\caption{Resolution tests: The circular velocity V$_c$= $\sqrt{M(<R)/R}$
 for run MW1hr (solid) and MW1g4 (dots). The mass distribution converges
 at radii larger than r $>$ 2.2 R$_d$ (vertical line).  The high
 resolution simulation contains significantly less baryons within a few times
 $\epsilon$}.
\label{fig:resolution}
\end{figure}

\begin{figure}
\begin{center}
\resizebox{7cm}{!}{\includegraphics{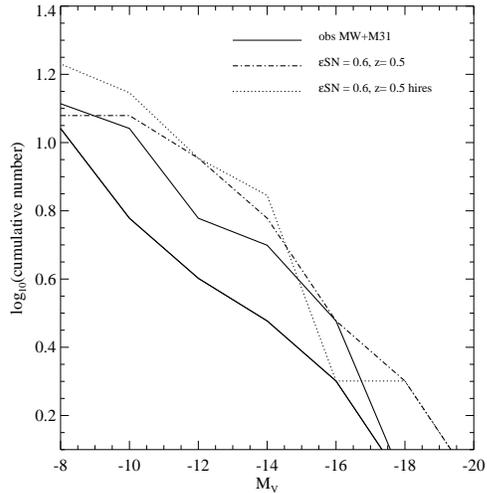}}
\end{center}
\caption{Resolution tests: The V -- band luminosity function of the
satellites system of MW1g4 (dashed) and its high resolution version
(dotted) at z = 0.5 compared with the Milky Way and Andromeda (solid
lines). }
\label{fig:Vband05}
\end{figure}

In this section, we show that our results are robust to resolution
effects. To do this, we have performed one additional simulation
(MW1hr) to  test if our results have converged as resolution is
further increased. A more complete study on how resolution affects the
formation of disks and the general properties of galaxies is highly
needed but beyond the scope of this paper. MW1hr uses the same initial
conditions as the MW1 runs, but uses eight times as many particles within
the high resolution region. $\epsilon$SN = 0.6 (as for run MW1g4) was
used.  Force resolution is improved by a factor of two to 0.3
kpc. Additional waves in the initial perturbations spectrum are
consistently added.  With a dark matter particle mass of only 7.6 $\times$
10$^5$ M$_\odot$ and star particle mass of 3 $\times$ 10$^4$ M$_\odot$ this
simulation has one of the highest resolutions for a Milky Way sized
galaxy in a cosmological context to date. Due to its high
computational cost we carried this simulation only to z $=$ 0.5, using
several tens of thousands of CPU hours.  Figure~\ref{fig:resolution}
compares the circular velocity V$_c$ of MW1g4 and MW1hr (again defined
as $\sqrt{M(r<R)/R}$) for the DM, baryonic and stellar components,
while Figure~\ref{fig:Vband05} compares the V--band satellites
luminosity function (again at z=0.5).

Figure~\ref{fig:resolution} shows that at the resolution of our
standard runs the total mass profile and the quantity of stars have
converged at radii corresponding to a couple of disk scale
lengths. V$_{rot}$ measured at 3.5R$_D$ (and used in Figures
12,~13,~17 and 18) is thus robust to resolution effects, at the
resolution of our standard runs. Test runs in a cosmological context
resolution (Governato et al in prep.) show a systematic increase in
V$_c$ and a decrease of stellar R$_D$ as resolution becomes
increasingly worse.  This result likely explains why our simulations
match the normalization of the TF relation. However at smaller radii
(only 3-4 times the gravitational softening) our standard run MW1g4
shows a much larger central mass than MW1hr.  This unphysical mass
concentration significantly raises the V$_c$ such that it peaks at
over 300km/sec. Because the standard and high resolution runs adopted
identical feedback and SF parameters, this result shows that
resolution plays an important role in the central region of
galaxies. In DM--only simulations, low resolution artificially lowers
central DM densities. In contrast with DM only simulations, low
resolution and lack of feedback in SPH simulations causes baryons to
become more centrally concentrated due to partially artificial angular momentum
loss. As resolution is increased, reduced artificial angular momentum
loss by both the stellar and gaseous component allow SN feedback to be
more effective and to reduce the central baryon concentration. Worse
resolution and possibly weaker feedback would have overestimated the
amount of baryons within 2.2 R$_d$.  As the force resolution is
increased the disk bar instability in the disk of MW1hr starts earlier
(z $\sim$ 2) and is more pronounced than in the MW1g4 simulation,
confirming studies by \citet{kaufmann06}. At this high resolution the
fraction of cold baryons is 40\% higher, and the mass in stars
is only 5\% lower. We also begin to resolve the thermal instabilities
predicted by \citet{maller04} and recently detected in simulations of
isolated galaxies \citep{kaufmann04}. It is encouraging that the
satellite V-band luminosity function (Fig.20) remains substantially
unaltered as resolution is increased, although a few more
satellites as faint as M$_V = -8$ are now resolved. This supports
results from S06 that our adopted SF+SN algorithm is relatively
insensitive to resolution.

\section{Conclusions}

We have presented a new set of simulations of the formation of disk
dominated galaxies in a full $\Lambda$CDM cosmological context carried
to the present time.  These simulations included a simple but
physically motivated recipe for star formation. We used isolated
galaxy models to calibrate the free parameters of our star
formation/feedback scheme to reproduce several important properties of
the stellar and ISM components of local disk galaxies over a range of
masses.  We then used this SF/feedback algorithm in cosmological
simulations with no further adjustment.

We simulated three different halos in a cosmological context, with
masses typical of those associated with disk galaxies.  These
cosmological simulations have a very high number of resolution
elements within the virial radius of the simulated galaxies ($\sim$
1.5 10$^5$ for each of the DM and gas components and a few hundred
thousand star particles), allowing us to resolve individual satellites
down to V$_c$ $\sim$ 20\% of the parent galaxy. A rotation dominated
stellar disk galaxy with a significant exponential component naturally
forms within each of these halos.  Our MW1 simulation provides a good
example of how the stellar disk can form shortly after a major merger
event, due to the large amount of orbital angular momentum transferred
to the gas component during the merger event.  Our three simulated
galaxies (DWF1, MW1, GAL1) share the size, stellar mass, colors and
current SFR typical of a low mass galaxy, of a Milky Way like galaxy
and a massive early type spiral, respectively.

One of our most significant results is that the simulated galaxies
fall on the TF and baryonic TF relations.  This match is possible in
part due to the high resolution of the simulations: Resolution is
crucial to avoid artificially losing the angular momentum gained this
way \citep{governato04,kaufmann06}. At our resolution, rotation curves
have been resolved down to a fraction of the stellar disk scale
lengths and (measured at 3.5 R$_d$) range from 135 to 270 km/sec in
amplitude.  The MW sized galaxy contains an amount of DM within the
solar radius well in agreement with observational constraints.  The
simulated galaxies also reproduced the observed 'anti-hierarchical'
trend of stellar populations: smaller galaxies have younger stellar
populations. As feedback delays star formation in protogalaxies a
larger reservoir of gas remains available to form disks until the bulk
of the mass of galaxy is assembled, typically between redshift 3 and 1.
Feedback and the cosmic UV field also drastically reduce the number of
galaxy satellites containing a significant stellar population.

We have complemented our standard runs with a high resolution test run
of our MW sized halo. This run has and unprecedented number of
particles: several million within R$_{vir}$. We used it to test the
convergence of results in our standard runs. At our highest resolution
the main galaxy is significantly less centrally concentrated, reducing
its bulge mass and its peak velocity considerably. While stellar
bulges are likely still unresolved in our standard runs, our tests and
estimates from previous works suggest that the mass distribution and
the rotation velocity of the cold gas and young stellar components
have converged at the scales of a few kpc. Such a high resolution coupled
with a physically motivated description of feedback explains why our
models do indeed fall onto the TF relations.

These simulations and those published in the recent literature show
 that some significant progress has been done toward understanding
 galaxy formation in the context of a $\Lambda$CDM concordance
 cosmology.  However, areas remain were progress needs to be made: (i)
 A larger range of initial conditions needs to be explored to
 understand the role played by halo spin and merging history on the
 morphology and colors of galaxies; (ii) Our resolution test shows
 that the inner mass distribution of the galaxies, including the
 bulge, is still poorly resolved by our standard runs and that
 possibly several million resolution elements per galaxy are
 needed to approach convergence; (iii) Satellites are still too
 bright compared to those orbiting our Milky Way.  While cosmic
 variance likely played a role (the largest satellites accreted by the
 MW1 system is more massive than the LMC), one obvious possibility to
 create fainter satellites would be to increase the efficiency of SN
 feedback, but that needs to be carefully calibrated with  the other
 constraints. Two intriguing possibilities that we plan to explore in
 future works are the increased feedback and UV field due to massive
 POPIII stars \citep{choudhuri05} and the local proximity effect that
 star formation from the main galaxy might have on the nearby dwarfs
 \citep{gnedin00}.  Overall, the possibility of obtaining a good match
 to the observed satellite LF seems within reach of future
 simulations.


\section*{Acknowledgments}

FG would like to thank A. Dekel. E.~D'Onghia, T.~Kaufmann, J.~Silk \&
M.~Steinmetz, for helpful conversations during this project. We thank
J.~Brinchmann, C.~Conselice, S.~McGaugh, L.~MacArthur and
R.~Giovanelli for sharing their data with us.  Support for this work,
part of the Spitzer Space Telescope Theoretical Research Program, was
provided by NASA through a contract issued by the Jet Propulsion
Laboratory, California Institute of Technology under a contract with
NASA. FG was a Brooks Fellow during the initial stages of this project
and was supported in part by NSF grants AST-0098557 and
AST-61-3931. TRQ acknowledges support from NSF grant AST-0098557 and
NSF grant PHY-0205413.  Most simulations were run at the Arctic Region
Supercomputing Center and on a local SGI multi-CPU machine.


\bibliographystyle{mn2e}
\bibliography{galgov4.5b.bib}

\begin{thebibliography}{}

\bibitem[\protect\citeauthoryear{{Abadi}, {Navarro}, {Steinmetz} \&
  {Eke}}{{Abadi} et~al.}{2003}]{abadi03a}
{Abadi} M.~G.,  {Navarro} J.~F.,  {Steinmetz} M.,    {Eke} V.~R.,  2003, \apj,
  591, 499

\bibitem[\protect\citeauthoryear{{Balogh}, {Pearce}, {Bower} \& {Kay}}{{Balogh}
  et~al.}{2001}]{balogh01}
{Balogh} M.~L.,  {Pearce} F.~R.,  {Bower} R.~G.,    {Kay} S.~T.,  2001, \mnras,
  326, 1228

\bibitem[\protect\citeauthoryear{{Balsara}}{{Balsara}}{1997}]{Balsara97}
{Balsara} D.~S.,  1997, in ASP Conf. Ser. 123: Computational Astrophysics; 12th
  Kingston Meeting {Modern Schemes for Solving Hyperbolic Conservation Laws of
  Interest in Computational Astrophysics on Parallel Machines}.
p.~274

\bibitem[\protect\citeauthoryear{{Bardeen}, {Bond}, {Kaiser} \&
  {Szalay}}{{Bardeen} et~al.}{1986}]{bbks}
{Bardeen} J.~M.,  {Bond} J.~R.,  {Kaiser} N.,    {Szalay} A.~S.,  1986, \apj,
  304, 15

\bibitem[\protect\citeauthoryear{{Barnes} \& {Hernquist}}{{Barnes} \&
  {Hernquist}}{1996}]{barneshernquist96}
{Barnes} J.~E.,  {Hernquist} L.,  1996, \apj, 471, 115

\bibitem[\protect\citeauthoryear{{Baugh}, {Cole} \& {Frenk}}{{Baugh}
  et~al.}{1996}]{baugh96}
{Baugh} C.~M.,  {Cole} S.,    {Frenk} C.~S.,  1996, \mnras, 283, 1361

\bibitem[\protect\citeauthoryear{{Bell} \& {de Jong}}{{Bell} \& {de
  Jong}}{2001}]{bell01}
{Bell} E.~F.,  {de Jong} R.~S.,  2001, \apj, 550, 212

\bibitem[\protect\citeauthoryear{{Benson}, {Frenk}, {Lacey}, {Baugh} \&
  {Cole}}{{Benson} et~al.}{2002}]{benson02}
{Benson} A.~J.,  {Frenk} C.~S.,  {Lacey} C.~G.,  {Baugh} C.~M.,    {Cole} S.,
  2002, \mnras, 333, 177

\bibitem[\protect\citeauthoryear{{Benz}}{{Benz}}{1990}]{benz90}
{Benz} W.,  1990, in Numerical Modelling of Nonlinear Stellar Pulsations
  Problems and Prospects {Smooth Particle Hydrodynamics - a Review}.
pp 269--+

\bibitem[\protect\citeauthoryear{{Binney}, {Gerhard} \& {Silk}}{{Binney}
  et~al.}{2001}]{binney01}
{Binney} J.,  {Gerhard} O.,    {Silk} J.,  2001, \mnras, 321, 471

\bibitem[\protect\citeauthoryear{{Bizyaev} \& {Mitronova}}{{Bizyaev} \&
  {Mitronova}}{2002}]{bizyaev02}
{Bizyaev} D.,  {Mitronova} S.,  2002, \aap, 389, 795

\bibitem[\protect\citeauthoryear{{Blanton} M.~R.}{{Blanton}}{2003}]{blanton03}
{Blanton} M.~R. e.~a.,  2003, \apj, 594, 186

\bibitem[\protect\citeauthoryear{{Blumenthal}, {Faber}, {Primack} \&
  {Rees}}{{Blumenthal} et~al.}{1984}]{blumenthal84}
{Blumenthal} G.~R.,  {Faber} S.~M.,  {Primack} J.~R.,    {Rees} M.~J.,  1984,
  \nat, 311, 517

\bibitem[\protect\citeauthoryear{{Brinchmann}, {Charlot}, {White}, {Tremonti},
  {Kauffmann}, {Heckman} \& {Brinkmann}}{{Brinchmann}
  et~al.}{2004}]{brinchmann04}
{Brinchmann} J.,  {Charlot} S.,  {White} S.~D.~M.,  {Tremonti} C.,  {Kauffmann}
  G.,  {Heckman} T.,    {Brinkmann} J.,  2004, \mnras, 351, 1151

\bibitem[\protect\citeauthoryear{{Broeils} \& {Rhee}}{{Broeils} \&
  {Rhee}}{1997}]{broeils97}
{Broeils} A.~H.,  {Rhee} M.-H.,  1997, \aap, 324, 877

\bibitem[\protect\citeauthoryear{{Brook}, {Kawata}, {Gibson} \&
  {Freeman}}{{Brook} et~al.}{2004}]{brook04}
{Brook} C.~B.,  {Kawata} D.,  {Gibson} B.~K.,    {Freeman} K.~C.,  2004, \apj,
  612, 894

\bibitem[\protect\citeauthoryear{{Bundy}, {Ellis} \& {Conselice}}{{Bundy}
  et~al.}{2005}]{bundy05}
{Bundy} K.,  {Ellis} R.,    {Conselice} C.,  2005, astro--ph/0502204

\bibitem[\protect\citeauthoryear{{Catinella}, {Giovanelli} \&
  {Haynes}}{{Catinella} et~al.}{2006}]{catinella06}
{Catinella} B.,  {Giovanelli} R.,    {Haynes} M.~P.,  2006, \apj, 640, 751

\bibitem[\protect\citeauthoryear{{Cen} \& {Ostriker}}{{Cen} \&
  {Ostriker}}{1999}]{cen99}
{Cen} R.,  {Ostriker} J.~P.,  1999, \apj, 514, 1

\bibitem[\protect\citeauthoryear{{Choudhury} \& {Ferrara}}{{Choudhury} \&
  {Ferrara}}{2005}]{choudhuri05}
{Choudhury} T.~R.,  {Ferrara} A.,  2005, \mnras, 361, 577

\bibitem[\protect\citeauthoryear{{Cole}, {Lacey}, {Baugh} \& {Frenk}}{{Cole}
  et~al.}{2000}]{cole00}
{Cole} S.,  {Lacey} C.~G.,  {Baugh} C.~M.,    {Frenk} C.~S.,  2000, \mnras,
  319, 168

\bibitem[\protect\citeauthoryear{{Conselice}, {Bundy}, {Ellis}, {Brichmann},
  {Vogt} \& {Phillips}}{{Conselice} et~al.}{2005}]{conselice05}
{Conselice} C.~J.,  {Bundy} K.,  {Ellis} R.~S.,  {Brichmann} J.,  {Vogt} N.~P.,
     {Phillips} A.~C.,  2005, \apj, 628, 160

\bibitem[\protect\citeauthoryear{{Cowie}, {Songaila}, {Hu} \& {Cohen}}{{Cowie}
  et~al.}{1996}]{cowie96}
{Cowie} L.~L.,  {Songaila} A.,  {Hu} E.~M.,    {Cohen} J.~G.,  1996, \aj, 112,
  839

\bibitem[\protect\citeauthoryear{{Davis}, {Efstathiou}, {Frenk} \&
  {White}}{{Davis} et~al.}{1985}]{davis85}
{Davis} M.,  {Efstathiou} G.,  {Frenk} C.~S.,    {White} S.~D.~M.,  1985, \apj,
  292, 371

\bibitem[\protect\citeauthoryear{{Debattista}, {Carollo}, {Mayer} \&
  {Moore}}{{Debattista} et~al.}{2004}]{debattista04}
{Debattista} V.~P.,  {Carollo} C.~M.,  {Mayer} L.,    {Moore} B.,  2004, \apjl,
  604, L93

\bibitem[\protect\citeauthoryear{{Debattista}, {Mayer}, {Carollo}, {Moore},
  {Wadsley} \& {Quinn}}{{Debattista} et~al.}{2005}]{debattista05}
{Debattista} V.~P.,  {Mayer} L.,  {Carollo} C.~M.,  {Moore} B.,  {Wadsley} J.,
    {Quinn} T.,  2005, astro-ph/0509310

\bibitem[\protect\citeauthoryear{{Dekel} \& {Woo}}{{Dekel} \&
  {Woo}}{2003}]{dekel03}
{Dekel} A.,  {Woo} J.,  2003, \mnras, 344, 1131

\bibitem[\protect\citeauthoryear{{Di Matteo}, {Springel} \& {Hernquist}}{{Di
  Matteo} et~al.}{2005}]{dimatteo05}
{Di Matteo} T.,  {Springel} V.,    {Hernquist} L.,  2005, \nat, 433, 604

\bibitem[\protect\citeauthoryear{{Diemand}, {Moore}, {Stadel} \&
  {Kazantzidis}}{{Diemand} et~al.}{2004}]{diemand04}
{Diemand} J.,  {Moore} B.,  {Stadel} J.,    {Kazantzidis} S.,  2004, \mnras,
  348, 977

\bibitem[\protect\citeauthoryear{{D'Onghia} E.}{{D'Onghia}}{2006}]{donghia06}
{D'Onghia} E. e.~a.,  2006, astro-ph/0602005

\bibitem[\protect\citeauthoryear{{Efstathiou}}{{Efstathiou}}{2000}]{efstathiou%
00}
{Efstathiou} G.,  2000, \mnras, 317, 697

\bibitem[\protect\citeauthoryear{{Efstathiou}, {Lake} \&
  {Negroponte}}{{Efstathiou} et~al.}{1982}]{efstathiou82}
{Efstathiou} G.,  {Lake} G.,    {Negroponte} J.,  1982, \mnras, 199, 1069

\bibitem[\protect\citeauthoryear{{Eke}, {Efstathiou} \& {Wright}}{{Eke}
  et~al.}{2000}]{eke00}
{Eke} V.,  {Efstathiou} G.,    {Wright} L.,  2000, \mnras, 315, L18

\bibitem[\protect\citeauthoryear{{Eke}, {Navarro} \& {Steinmetz}}{{Eke}
  et~al.}{2001}]{eke01}
{Eke} V.~R.,  {Navarro} J.~F.,    {Steinmetz} M.,  2001, \apj, 554, 114

\bibitem[\protect\citeauthoryear{{Elmegreen}, {Elmegreen}, {Vollbach}, {Foster}
  \& {Ferguson}}{{Elmegreen} et~al.}{2005}]{elmegreen05}
{Elmegreen} B.~G.,  {Elmegreen} D.~M.,  {Vollbach} D.~R.,  {Foster} E.~R.,
  {Ferguson} T.~E.,  2005, \apj, 634, 101

\bibitem[\protect\citeauthoryear{{Evrard}}{{Evrard}}{1988}]{evrard88}
{Evrard} A.~E.,  1988, \mnras, 235, 911

\bibitem[\protect\citeauthoryear{{Fall}}{{Fall}}{1983}]{fall83}
{Fall} S.~M.,  1983, in IAU Symp. 100: Internal Kinematics and Dynamics of
  Galaxies {Galaxy formation - Some comparisons between theory and
  observation}.
pp 391--398

\bibitem[\protect\citeauthoryear{{Fall} \& {Efstathiou}}{{Fall} \&
  {Efstathiou}}{1980}]{fall80}
{Fall} S.~M.,  {Efstathiou} G.,  1980, \mnras, 193, 189

\bibitem[\protect\citeauthoryear{{Ferrara}}{{Ferrara}}{2002}]{ferrara02}
{Ferrara} A.,  2002, in ASP Conf. Ser. 285: Modes of Star Formation and the
  Origin of Field Populations {Stellar Feedback, Dark Matter and Dwarf
  Evolution (Review)}

\bibitem[\protect\citeauthoryear{{Ferreras}, {Silk}, {B{\" o}hm} \&
  {Ziegler}}{{Ferreras} et~al.}{2004}]{ferreras05}
{Ferreras} I.,  {Silk} J.,  {B{\" o}hm} A.,    {Ziegler} B.,  2004, \mnras,
  355, 64

\bibitem[\protect\citeauthoryear{{Florido}, {Battaner}, {Guijarro},
  {Garz{\'o}n} \& {Jim{\'e}nez-Vicente}}{{Florido} et~al.}{2001}]{florido01}
{Florido} E.,  {Battaner} E.,  {Guijarro} A.,  {Garz{\'o}n} F.,
  {Jim{\'e}nez-Vicente} J.,  2001, \aap, 378, 82

\bibitem[\protect\citeauthoryear{{Frenk}, {White}, {Efstathiou} \&
  {Davis}}{{Frenk} et~al.}{1985}]{frenk85}
{Frenk} C.~S.,  {White} S.~D.~M.,  {Efstathiou} G.,    {Davis} M.,  1985, \nat,
  317, 595

\bibitem[\protect\citeauthoryear{{Gallazzi}, {Charlot}, {Brinchmann}, {White}
  \& {Tremonti}}{{Gallazzi} et~al.}{2005}]{gallazzi05}
{Gallazzi} A.,  {Charlot} S.,  {Brinchmann} J.,  {White} S.~D.~M.,
  {Tremonti} C.~A.,  2005, \mnras, 362, 41

\bibitem[\protect\citeauthoryear{{Gardner}}{{Gardner}}{2001}]{gardner01}
{Gardner} J.~P.,  2001, \apj, 557, 616

\bibitem[\protect\citeauthoryear{{Gerritsen}}{{Gerritsen}}{1997}]{gerritsen97}
{Gerritsen} J.~P.~E.,  1997, Ph.D.~Thesis

\bibitem[\protect\citeauthoryear{{Giovanelli}, {Haynes}, {Herter}, {Vogt}, {da
  Costa}, {Freudling}, {Salzer} \& {Wegner}}{{Giovanelli}
  et~al.}{1997}]{giovanelli97a}
{Giovanelli} R.,  {Haynes} M.~P.,  {Herter} T.,  {Vogt} N.~P.,  {da Costa}
  L.~N.,  {Freudling} W.,  {Salzer} J.~J.,    {Wegner} G.,  1997, \aj, 113, 53

\bibitem[\protect\citeauthoryear{{Gnedin}}{{Gnedin}}{2000}]{gnedin00}
{Gnedin} N.~Y.,  2000, \apj, 535, 530

\bibitem[\protect\citeauthoryear{{Governato}, {Mayer}, {Wadsley}, {Gardner},
  {Willman}, {Hayashi}, {Quinn}, {Stadel} \& {Lake}}{{Governato}
  et~al.}{2004}]{governato04}
{Governato} F.,  {Mayer} L.,  {Wadsley} J.,  {Gardner} J.~P.,  {Willman} B.,
  {Hayashi} E.,  {Quinn} T.,  {Stadel} J.,    {Lake} G.,  2004, \apj, 607, 688

\bibitem[\protect\citeauthoryear{{Governato}, {Moore}, {Cen}, {Stadel}, {Lake}
  \& {Quinn}}{{Governato} et~al.}{1997}]{governato97}
{Governato} F.,  {Moore} B.,  {Cen} R.,  {Stadel} J.,  {Lake} G.,    {Quinn}
  T.,  1997, New Astronomy, 2, 91

\bibitem[\protect\citeauthoryear{{Granato}, {De Zotti}, {Silva}, {Bressan} \&
  {Danese}}{{Granato} et~al.}{2004}]{granato04}
{Granato} G.~L.,  {De Zotti} G.,  {Silva} L.,  {Bressan} A.,    {Danese} L.,
  2004, \apj, 600, 580

\bibitem[\protect\citeauthoryear{{Haardt} \& {Madau}}{{Haardt} \&
  {Madau}}{1996}]{haardtmadau96}
{Haardt} F.,  {Madau} P.,  1996, \apj, 461, 20

\bibitem[\protect\citeauthoryear{{Haehnelt}, {Steinmetz} \& {Rauch}}{{Haehnelt}
  et~al.}{1998}]{haehnelt98}
{Haehnelt} M.~G.,  {Steinmetz} M.,    {Rauch} M.,  1998, \apj, 495, 647

\bibitem[\protect\citeauthoryear{{Hopkins}, {Hernquist}, {Cox}, {Di Matteo},
  {Martini}, {Robertson} \& {Springel}}{{Hopkins} et~al.}{2005}]{hopkins05}
{Hopkins} P.~F.,  {Hernquist} L.,  {Cox} T.~J.,  {Di Matteo} T.,  {Martini} P.,
   {Robertson} B.,    {Springel} V.,  2005, \apj, 630, 705

\bibitem[\protect\citeauthoryear{{Katz}}{{Katz}}{1992}]{katz92}
{Katz} N.,  1992, \apj, 391, 502

\bibitem[\protect\citeauthoryear{{Katz}, {Weinberg} \& {Hernquist}}{{Katz}
  et~al.}{1996}]{katz96}
{Katz} N.,  {Weinberg} D.~H.,    {Hernquist} L.,  1996, \apjs, 105, 19

\bibitem[\protect\citeauthoryear{{Kaufmann}, {Mayer}, {Moore}, {Stadel} \&
  {Wadsley}}{{Kaufmann} et~al.}{2004}]{kaufmann04}
{Kaufmann} T.,  {Mayer} L.,  {Moore} B.,  {Stadel} J.,    {Wadsley} J.,  2004,
  Baryons in Dark Matter Halos

\bibitem[\protect\citeauthoryear{{Kaufmann} \& {Mayer}}{{Kaufmann} \&
  {Mayer}}{2006}]{kaufmann06}
{Kaufmann} T.,  {Mayer} L. e.~a.,  2006, ArXiv Astrophysics e-prints

\bibitem[\protect\citeauthoryear{{Kawata}, {Arimoto}, {Cen} \&
  {Gibson}}{{Kawata} et~al.}{2005}]{kawata06}
{Kawata} D.,  {Arimoto} N.,  {Cen} R.,    {Gibson} B.,  2005, astro-ph/0509402

\bibitem[\protect\citeauthoryear{{Kennicutt}}{{Kennicutt}}{1998}]{kennicutt98}
{Kennicutt} R.~C.,  1998, \apj, 498, 541

\bibitem[\protect\citeauthoryear{{Klypin}, {Kravtsov}, {Valenzuela} \&
  {Prada}}{{Klypin} et~al.}{1999}]{klypin99}
{Klypin} A.,  {Kravtsov} A.~V.,  {Valenzuela} O.,    {Prada} F.,  1999, \apj,
  522, 82

\bibitem[\protect\citeauthoryear{{Klypin}, {Zhao} \& {Somerville}}{{Klypin}
  et~al.}{2002}]{klypin02}
{Klypin} A.,  {Zhao} H.,    {Somerville} R.~S.,  2002, \apj, 573, 597

\bibitem[\protect\citeauthoryear{{Kormendy} \& {Kennicutt}}{{Kormendy} \&
  {Kennicutt}}{2004}]{kormendy04}
{Kormendy} J.,  {Kennicutt} R.~C.,  2004, 42, 603

\bibitem[\protect\citeauthoryear{{Kravtsov} \& {Gnedin}}{{Kravtsov} \&
  {Gnedin}}{2005}]{kravtsov05}
{Kravtsov} A.~V.,  {Gnedin} O.~Y.,  2005, \apj, 623, 650

\bibitem[\protect\citeauthoryear{{Kravtsov}, {Gnedin} \& {Klypin}}{{Kravtsov}
  et~al.}{2004}]{kravtsov04}
{Kravtsov} A.~V.,  {Gnedin} O.~Y.,    {Klypin} A.~A.,  2004, \apj, 609, 482

\bibitem[\protect\citeauthoryear{{Krumholz} \& {McKee}}{{Krumholz} \&
  {McKee}}{2005}]{krumholz05}
{Krumholz} M.,  {McKee} C.,  2005, astro-ph/0505177

\bibitem[\protect\citeauthoryear{{Lake} \& {Carlberg}}{{Lake} \&
  {Carlberg}}{1988}]{lakecarlberg88}
{Lake} G.,  {Carlberg} R.~G.,  1988, \aj, 96, 1587

\bibitem[\protect\citeauthoryear{{Li}, {Mo} \& {van den Bosch}}{{Li}
  et~al.}{2005}]{li05}
{Li} Y.,  {Mo} H.,    {van den Bosch} F.,  2005

\bibitem[\protect\citeauthoryear{{Mac Low} \& {Ferrara}}{{Mac Low} \&
  {Ferrara}}{1999}]{maclow99}
{Mac Low} M.,  {Ferrara} A.,  1999, \apj, 513, 142

\bibitem[\protect\citeauthoryear{{MacArthur}, {Courteau}, {Bell} \&
  {Holtzman}}{{MacArthur} et~al.}{2004}]{macarthur04}
{MacArthur} L.~A.,  {Courteau} S.,  {Bell} E.,    {Holtzman} J.~A.,  2004,
  \apjs, 152, 175

\bibitem[\protect\citeauthoryear{{Macci{\` o}}, {Governato} \&
  {Horellou}}{{Macci{\` o}} et~al.}{2005}]{maccio05}
{Macci{\` o}} A.~V.,  {Governato} F.,    {Horellou} C.,  2005, \mnras, 359, 941

\bibitem[\protect\citeauthoryear{{Maller} \& {Bullock}}{{Maller} \&
  {Bullock}}{2004}]{maller04}
{Maller} A.~H.,  {Bullock} J.~S.,  2004, \mnras, 355, 694

\bibitem[\protect\citeauthoryear{{Maller} \& {Dekel}}{{Maller} \&
  {Dekel}}{2002}]{maller02}
{Maller} A.~H.,  {Dekel} A.,  2002, \mnras, 335, 487

\bibitem[\protect\citeauthoryear{{Martin}}{{Martin}}{1999}]{martin99}
{Martin} C.~L.,  1999, \apj, 513, 156

\bibitem[\protect\citeauthoryear{{Martin}, {Fanson} \& {Schiminovich}}{{Martin}
  et~al.}{2005}]{martin05}
{Martin} D.~C.,  {Fanson} J.,    {Schiminovich} D. e.~a.,  2005, \apjl, 619, L1

\bibitem[\protect\citeauthoryear{{Mateo}}{{Mateo}}{1998}]{mateo98}
{Mateo} M.~L.,  1998, 36, 435

\bibitem[\protect\citeauthoryear{{Mayer} \& {Moore}}{{Mayer} \&
  {Moore}}{2004}]{mayer04b}
{Mayer} L.,  {Moore} B.,  2004, \mnras, 354, 477

\bibitem[\protect\citeauthoryear{{Mayer} \& {Wadsley}}{{Mayer} \&
  {Wadsley}}{2004}]{mayer04a}
{Mayer} L.,  {Wadsley} J.,  2004, \mnras, 347, 277

\bibitem[\protect\citeauthoryear{{McGaugh}}{{McGaugh}}{2005}]{mcgaugh05}
{McGaugh} S.~S.,  2005, \apj, 632, 859

\bibitem[\protect\citeauthoryear{{McGaugh}, {Schombert}, {Bothun} \& {de
  Blok}}{{McGaugh} et~al.}{2000}]{mcgaugh00}
{McGaugh} S.~S.,  {Schombert} J.~M.,  {Bothun} G.~D.,    {de Blok} W.~J.~G.,
  2000, \apjl, 533, L99

\bibitem[\protect\citeauthoryear{{McKee} \& {Ostriker}}{{McKee} \&
  {Ostriker}}{1977}]{mckee77}
{McKee} C.~F.,  {Ostriker} J.~P.,  1977, \apj, 218, 148

\bibitem[\protect\citeauthoryear{{Mo} \& {Miralda-Escude}}{{Mo} \&
  {Miralda-Escude}}{1996}]{mo96}
{Mo} H.~J.,  {Miralda-Escude} J.,  1996, \apj, 469, 589

\bibitem[\protect\citeauthoryear{{Monaco}}{{Monaco}}{2004}]{monaco04}
{Monaco} P.,  2004, \mnras, 352, 181

\bibitem[\protect\citeauthoryear{{Monaghan}}{{Monaghan}}{1992}]{monaghan92}
{Monaghan} J.~J.,  1992, 30, 543

\bibitem[\protect\citeauthoryear{{Moore}, {Ghigna}, {Governato}, {Lake},
  {Quinn}, {Stadel} \& {Tozzi}}{{Moore} et~al.}{1999}]{moore99}
{Moore} B.,  {Ghigna} S.,  {Governato} F.,  {Lake} G.,  {Quinn} T.,  {Stadel}
  J.,    {Tozzi} P.,  1999, \apjl, 524, L19

\bibitem[\protect\citeauthoryear{{Navarro} \& {Steinmetz}}{{Navarro} \&
  {Steinmetz}}{1997}]{navarrosteinmetz97}
{Navarro} J.~F.,  {Steinmetz} M.,  1997, \apj, 478, 13

\bibitem[\protect\citeauthoryear{{Navarro} \& {Steinmetz}}{{Navarro} \&
  {Steinmetz}}{2000}]{navarrosteinmetz00}
{Navarro} J.~F.,  {Steinmetz} M.,  2000, \apj, 538, 477

\bibitem[\protect\citeauthoryear{{Navarro} \& {White}}{{Navarro} \&
  {White}}{1994}]{navarro94}
{Navarro} J.~F.,  {White} S.~D.~M.,  1994, \mnras, 267, 401

\bibitem[\protect\citeauthoryear{{Neistein}, {van den Bosch} \&
  {Dekel}}{{Neistein} et~al.}{2006}]{neinstein06}
{Neistein} E.,  {van den Bosch} F.~C.,    {Dekel} A.,  2006, \mnras, 372, 933

\bibitem[\protect\citeauthoryear{{Nordstr{\" o}m}, {Mayor}, {Andersen},
  {Holmberg}, {Pont}, {J{\o}rgensen}, {Olsen}, {Udry} \& {Mowlavi}}{{Nordstr{\"
  o}m} et~al.}{2004}]{nordstrom04}
{Nordstr{\" o}m} B.,  {Mayor} M.,  {Andersen} J.,  {Holmberg} J.,  {Pont} F.,
  {J{\o}rgensen} B.~R.,  {Olsen} E.~H.,  {Udry} S.,    {Mowlavi} N.,  2004,
  \aap, 418, 989

\bibitem[\protect\citeauthoryear{{Okamoto}, {Eke}, {Frenk} \&
  {Jenkins}}{{Okamoto} et~al.}{2005}]{okamoto05}
{Okamoto} T.,  {Eke} V.~R.,  {Frenk} C.~S.,    {Jenkins} A.,  2005, \mnras,
  363, 1299

\bibitem[\protect\citeauthoryear{{Ostriker} \& {McKee}}{{Ostriker} \&
  {McKee}}{1988}]{ostrikermckee88}
{Ostriker} J.~P.,  {McKee} C.~F.,  1988, Reviews of Modern Physics, 60, 1

\bibitem[\protect\citeauthoryear{{Peebles} \& {Ratra}}{{Peebles} \&
  {Ratra}}{2003}]{peeblesratra03}
{Peebles} P.~J.,  {Ratra} B.,  2003, Reviews of Modern Physics, 75, 559

\bibitem[\protect\citeauthoryear{{Peebles}}{{Peebles}}{1969}]{peebles69}
{Peebles} P.~J.~E.,  1969, \apj, 155, 393

\bibitem[\protect\citeauthoryear{{Pizzella}, {Corsini}, {Vega Beltr{\' a}n} \&
  {Bertola}}{{Pizzella} et~al.}{2004}]{pizzella04}
{Pizzella} A.,  {Corsini} E.~M.,  {Vega Beltr{\' a}n} J.~C.,    {Bertola} F.,
  2004, \aap, 424, 447

\bibitem[\protect\citeauthoryear{{Pohlen}, {Balcells}, {L{\"u}tticke} \&
  {Dettmar}}{{Pohlen} et~al.}{2004}]{pohlen04s0}
{Pohlen} M.,  {Balcells} M.,  {L{\"u}tticke} R.,    {Dettmar} R.-J.,  2004,
  \aap, 422, 465

\bibitem[\protect\citeauthoryear{{Pohlen} \& {Trujillo}}{{Pohlen} \&
  {Trujillo}}{2005}]{pohlen05}
{Pohlen} M.,  {Trujillo} I.,  2005

\bibitem[\protect\citeauthoryear{{Quinn}, {Katz} \& {Efstathiou}}{{Quinn}
  et~al.}{1996}]{quinn96}
{Quinn} T.,  {Katz} N.,    {Efstathiou} G.,  1996, \mnras, 278, L49

\bibitem[\protect\citeauthoryear{{Raiteri}, {Villata} \& {Navarro}}{{Raiteri}
  et~al.}{1996}]{raiteri96}
{Raiteri} C.~M.,  {Villata} M.,    {Navarro} J.~F.,  1996, \aap, 315, 105

\bibitem[\protect\citeauthoryear{{Robertson}, {Yoshida}, {Springel} \&
  {Hernquist}}{{Robertson} et~al.}{2004}]{robertson04}
{Robertson} B.,  {Yoshida} N.,  {Springel} V.,    {Hernquist} L.,  2004, \apj,
  606, 32

\bibitem[\protect\citeauthoryear{{Ryder} \& {Dopita}}{{Ryder} \&
  {Dopita}}{1994}]{ryder94}
{Ryder} S.~D.,  {Dopita} M.~A.,  1994, \apj, 430, 142

\bibitem[\protect\citeauthoryear{{Seth}, {Dalcanton} \& {de Jong}}{{Seth}
  et~al.}{2005}]{seth05}
{Seth} A.~C.,  {Dalcanton} J.~J.,    {de Jong} R.~S.,  2005, \aj, 129, 1331

\bibitem[\protect\citeauthoryear{{Shen}, {Mo}, {White}, {Blanton}, {Kauffmann},
  {Voges}, {Brinkmann} \& {Csabai}}{{Shen} et~al.}{2003}]{shen03}
{Shen} S.,  {Mo} H.~J.,  {White} S.~D.~M.,  {Blanton} M.~R.,  {Kauffmann} G.,
  {Voges} W.,  {Brinkmann} J.,    {Csabai} I.,  2003, \mnras, 343, 978

\bibitem[\protect\citeauthoryear{{Silk}}{{Silk}}{2001}]{silk01}
{Silk} J.,  2001, \mnras, 324, 313

\bibitem[\protect\citeauthoryear{{Silva}, {Granato}, {Bressan} \&
  {Danese}}{{Silva} et~al.}{1998}]{silva98}
{Silva} L.,  {Granato} G.~L.,  {Bressan} A.,    {Danese} L.,  1998, \apj, 509,
  103

\bibitem[\protect\citeauthoryear{{Slyz}, {Devriendt}, {Bryan} \& {Silk}}{{Slyz}
  et~al.}{2005}]{Slyz05}
{Slyz} A.~D.,  {Devriendt} J.~E.~G.,  {Bryan} G.,    {Silk} J.,  2005, \mnras,
  356, 737

\bibitem[\protect\citeauthoryear{{Sommer-Larsen}, {G{\" o}tz} \&
  {Portinari}}{{Sommer-Larsen} et~al.}{2003}]{sommer03}
{Sommer-Larsen} J.,  {G{\" o}tz} M.,    {Portinari} L.,  2003, \apj, 596, 47

\bibitem[\protect\citeauthoryear{{Spergel}, {Verde}, {Peiris}, {Komatsu},
  {Nolta}, {Bennett}, {Halpern}, {Hinshaw}, {Jarosik}, {Kogut}, {Limon},
  {Meyer}, {Page}, {Tucker}, {Weiland}, {Wollack} \& {Wright}}{{Spergel}
  et~al.}{2003}]{spergel03}
{Spergel} D.~N.,  {Verde} L.,  {Peiris} H.~V.,  {Komatsu} E.,  {Nolta} M.~R.,
  {Bennett} C.~L.,  {Halpern} M.,  {Hinshaw} G.,  {Jarosik} N.,  {Kogut} A.,
  {Limon} M.,  {Meyer} S.~S.,  {Page} L.,  {Tucker} G.~S.,  {Weiland} J.~L.,
  {Wollack} E.,    {Wright} E.~L.,  2003, \apjs, 148, 175

\bibitem[\protect\citeauthoryear{{Springel}}{{Springel}}{2000}]{springel00}
{Springel} V.,  2000, \mnras, 312, 859

\bibitem[\protect\citeauthoryear{{Springel} \& {Hernquist}}{{Springel} \&
  {Hernquist}}{2002}]{springel02}
{Springel} V.,  {Hernquist} L.,  2002, \mnras, 333, 649

\bibitem[\protect\citeauthoryear{{Springel} \& {Hernquist}}{{Springel} \&
  {Hernquist}}{2005}]{springelhernquist05}
{Springel} V.,  {Hernquist} L.,  2005, \apjl, 622, L9

\bibitem[\protect\citeauthoryear{{Steinmetz} \& {Navarro}}{{Steinmetz} \&
  {Navarro}}{2002}]{steinmetznavarro02}
{Steinmetz} M.,  {Navarro} J.~F.,  2002, New Astronomy, 7, 155

\bibitem[\protect\citeauthoryear{{Stinson}, {Seth}, {Katz}, {Governato},
  {Wadsley} \& {Quinn}}{{Stinson} et~al.}{2006}]{stinson06}
{Stinson} G.,  {Seth} A.,  {Katz} N.,  {Governato} F.,  {Wadsley} J.,
  {Quinn} T.,  2006, astro-ph/0602350

\bibitem[\protect\citeauthoryear{{Tasker} \& {Bryan}}{{Tasker} \&
  {Bryan}}{2005}]{tasker05}
{Tasker} E.~J.,  {Bryan} G.,  {2005}

\bibitem[\protect\citeauthoryear{{Thacker} \& {Couchman}}{{Thacker} \&
  {Couchman}}{2000}]{thacker00}
{Thacker} R.~J.,  {Couchman} H.~M.~P.,  2000, \apj, 545, 728

\bibitem[\protect\citeauthoryear{{Tittley}, {Pearce} \& {Couchman}}{{Tittley}
  et~al.}{2001}]{tittley01}
{Tittley} E.~R.,  {Pearce} F.~R.,    {Couchman} H.~M.~P.,  2001, \apj, 561, 69

\bibitem[\protect\citeauthoryear{{Toomre} \& {Toomre}}{{Toomre} \&
  {Toomre}}{1972}]{toomre72}
{Toomre} A.,  {Toomre} J.,  1972, \apj, 178, 623

\bibitem[\protect\citeauthoryear{Trujillo et~al.,}{Trujillo
  et~al.}{2005}]{trujillo05}
Trujillo I.,  et~al., {2005}, astro-ph/0504255

\bibitem[\protect\citeauthoryear{{Valenzuela} \& {Klypin}}{{Valenzuela} \&
  {Klypin}}{2003}]{valenzuela03}
{Valenzuela} O.,  {Klypin} A.,  2003, \mnras, 345, 406

\bibitem[\protect\citeauthoryear{{Wada} \& {Norman}}{{Wada} \&
  {Norman}}{2001}]{wada01}
{Wada} K.,  {Norman} C.~A.,  2001, \apj, 547, 172

\bibitem[\protect\citeauthoryear{{Wadsley}, {Stadel} \& {Quinn}}{{Wadsley}
  et~al.}{2004}]{wadsley04}
{Wadsley} J.~W.,  {Stadel} J.,    {Quinn} T.,  2004, New Astronomy, 9, 137

\bibitem[\protect\citeauthoryear{{West}, {Garcia-Appadoo}, {Dalcanton} \&
  {Disney}}{{West} et~al.}{2005}]{west05}
{West} A.~A.,  {Garcia-Appadoo} D.~A.,  {Dalcanton} J.~J.,    {Disney} M.~J.,
  2005, in AIP Conf. Proc. 761: The Spectral Energy Distributions of Gas-Rich
  Galaxies: Confronting Models with Data {HI Selected Galaxies in the SDSS}.
pp 409--+

\bibitem[\protect\citeauthoryear{{White} \& {Frenk}}{{White} \&
  {Frenk}}{1991}]{whitefrenk91}
{White} S.~D.~M.,  {Frenk} C.~S.,  1991, \apj, 379, 52

\bibitem[\protect\citeauthoryear{{White} \& {Rees}}{{White} \&
  {Rees}}{1978}]{whiterees78}
{White} S.~D.~M.,  {Rees} M.~J.,  1978, \mnras, 183, 341

\bibitem[\protect\citeauthoryear{{Willman}, {Governato}, {Dalcanton}, {Reed} \&
  {Quinn}}{{Willman} et~al.}{2004}]{willman04}
{Willman} B.,  {Governato} F.,  {Dalcanton} J.~J.,  {Reed} D.,    {Quinn} T.,
  2004, \mnras, 353, 639

\bibitem[\protect\citeauthoryear{{Yepes}, {Kates}, {Khokhlov} \&
  {Klypin}}{{Yepes} et~al.}{1997}]{yepes97}
{Yepes} G.,  {Kates} R.,  {Khokhlov} A.,    {Klypin} A.,  1997, \mnras, 284,
  235

\bibitem[\protect\citeauthoryear{{Yoachim} \& {Dalcanton}}{{Yoachim} \&
  {Dalcanton}}{2005}]{yoachim05}
{Yoachim} P.,  {Dalcanton} J.~J.,  2005, \apj, 624, 701

\bibitem[\protect\citeauthoryear{{Yoachim} \& {Dalcanton}}{{Yoachim} \&
  {Dalcanton}}{2006}]{yoachim06}
{Yoachim} P.,  {Dalcanton} J.~J.,  2006, \aj, 131, 226

\end{thebibliography}



\end{document}